\documentclass[nofootinbib,12pt]{revtex4}%
\usepackage{amssymb}
\usepackage{amsmath}
\usepackage{amsfonts}
\usepackage{graphicx}%
\begin{document}
\title{Generalizations of Lunin-Maldacena transformation on the $AdS_{5}\times S^{5}$ background}
\author{R. C. Rashkov}
\email{rash@phys.uni-sofia.bg}
\affiliation{Department of Physics, Sofia University, 1164 Sofia, Bulgaria}
\author{K. S. Viswanathan}
\email{kviswana@sfu.ca}
\affiliation{Department of Physics, Simon Fraser University, Burnaby BC, Canada}
\author{Yi Yang}
\email{yiyang@mail.nctu.edu.tw}
\affiliation{Department of Electrophysics, National Chiao Tung University, Hsinchu, Taiwan.}
\date{\today}

\begin{abstract}
In this paper we consider a simple generalization of the method of Lunin and
Maldacena for generating new string backgrounds based on TsT-transformations.
We study multi-shift $Ts\cdots sT$ transformations applied to backgrounds with
at least two U(1) isometries. We prove that the string currents in any two
backgrounds related by Ts...sT-transformations are equal. Applying this
procedure to the $AdS_{5}\times S^{5}$, we find a new background and study
some properties of the semiclassical strings.

\end{abstract}
\maketitle
%

%TCIMACRO{\TeXButton{equation number}{\setcounter{equation}{0}
%\renewcommand{\theequation}{\arabic{section}.\arabic{equation}}}}%
%BeginExpansion
\setcounter{equation}{0}
\renewcommand{\theequation}{\arabic{section}.\arabic{equation}}%
%EndExpansion

\section{Introduction}

In this paper we present a simple generalization of the method for obtaining
deformed string backgrounds proposed by Lunin and Maldacena
\cite{hep-th/0502086} and developed in detail by Frolov \cite{hep-th/0503201}.
The method in the above papers is based on T-duality on one of $U(1)$
variables, shift of another $U(1)$ variable and T-duality back on the first
$U(1)$ variable (called TsT-transformation)\footnote{See the discussion in
Section II.}. Our method consists in multi shifts at the second step which
allows one to obtain new string backgrounds (we call this $Ts_{1}\cdots s_{n}
T$ transformation). We prove also that the $U(1)$ string currents in any two
backgrounds related by $Ts_{1}\cdots s_{n} T$ transformations are equal. We
present also an application of our method to string theory in $AdS_{5}\times
S^{5}$ background.

In the past few years the main efforts in string theory were directed towards
establishing string/gauge theory correspondence. The vast majority of papers
were on qualitative and quantitative description of $\mathcal{N}=4$ SYM theory
with $SU(N)$ gauge group by making use of string sigma model on $AdS_{5}\times
S^{5}$ \cite{hep-th/9711200,hep-th/9802150,hep-th/9802109}. The AdS/CFT
correspondence implies that the energy of closed string states is equal to the
anomalous dimensions of certain local SYM operators
\cite{hep-th/0202021,hep-th/0204051}. At supergravity level this
correspondence has been checked in a number of cases (for review see for
instance \cite{hep-th/0201253}) but the match between the string energy and
the anomalous dimensions beyond that approximation still remains a challenge.

The first important step in establishing AdS/CFT correspondence is to obtain
the spectrum of the anomalous dimensions of the primary operators made up of
local gauge fields. On string theory side it requires one not only to solve
the theory at classical level but also include its quantization.

The main challenge in quantizing string theory is that it is highly non-linear
and thus difficult to manage. The only option available so far is to look at
the semiclassical region of large quantum numbers where the results are
reliable. On gauge theory side the derivation of the anomalous dimensions is
also a difficult task. A breakthrough in this direction has been the
observation of Minahan and Zarembo that one loop dilatation operator
restricted to the bosonic sector of N=4 SYM theory can be interpreted as the
Hamiltonian of integrable spin chain \cite{hep-th/0212208}. This observation
raised the question about the dilatation operator in N=4 SYM theory and
integrability (for a recent review see for instance \cite{hep-th/0407277} and
references therein).

On the other hand, the question of reduction of the string sigma model to
particular integrable systems and the question of integrability of string
theory at classical and quantum level was considered in a number of papers
\cite{hep-th/0411089, hep-th/0311004, hep-th/0311203, hep-th/0311203,
hep-th/0403121}. The intensive study of "nearly" BPS, or BMN, quantum strings
and non-BPS ones give a remarkable match with the results from gauge theory
side at least at the first few loops
\cite{hep-th/0212208,hep-th/0411089,hep-th/0305116,hep-th/0407277,hep-th/0308117,hep-th/0311203,hep-th/0403120,hep-th/0405167,hep-th/0403139,hep-th/0502188,hep-th/0403121}%
. This match however is not a coincidence. In the above papers it was
suggested that certain spin chains should describe particular string sectors
and thus should allow the comparison to the gauge theory computations.
Subsequently it has been found that the match between string theory and SYM
theory in the examples discussed above lies in the Yangian symmetries
responsible to large extent for the integrability on both sides
\cite{hep-th/0401243, hep-th/0409180}. Since in this paper we will consider
the string theory side, we refer the reader to the above papers for details on
this connection.

>From the picture emerging from the above studies one can conclude that the
integrable structures play an important role in establishing the AdS/CFT
correspondence at classical and hopefully at quantum level as well.

Although we already have some understanding of string/gauge theory
correspondence in the case of $AdS_{5}\times S^{5}$ background and
$\mathcal{N}=4$ SYM, much less is known in the case of theories with less than
maximal supersymmetry. There have been some studies of AdS/CFT correspondence
for less supersymmetric string backgrounds
\cite{hep-th/0304035,hep-th/0410262,hep-th/0211197,hep-th/0404122,
hep-th/0309114,hep-th/0401058,hep-th/0210218,hep-th/0210122,
hep-th/0402007,hep-th/0302005,hep-th/0501072,hep-th/0408014,
hep-th/0403261,hep-th/0409205,hep-th/0301178,hep-th/0307113,
hep-th/0308087,hep-th/0309161,hep-th/0411023,hep-th/0212061,
hep-th/0209224,hep-th/0402202}. However it is not quite clear how exactly to
implement the correspondence. The main obstacles lie in knowing if and how
Kaluza-Klein modes naturally present in such backgrounds contribute to the
string energy, which corner in the space of gauge operators is described by
these strings, and are these sub sectors closed under the renormalization
group flow?

An important step towards a deeper understanding of AdS/CFT correspondence in
its less supersymmetric sector was recently given by Lunin and Maldacena
\cite{hep-th/0502086}. From gauge theory point of view the possible
deformations of $\mathcal{N}=4$ SYM gauge theory that break the supersymmetry
were studied by Leigh and Strassler \cite{hep-th/9503121}. It should be
mentioned that the deformations of $\mathcal{N}=4$ SYM theory and integrable
spin chains have been considered in some detail in \cite{hep-th/0405215}. In
\cite{hep-th/0502086} Lunin and Maldacena found the gravity dual to the
$\beta$-deformations of $\mathcal{N}=4$ SYM theory studied in
\cite{hep-th/9503121}. They demonstrated that a certain deformation of the
$AdS_{5}\times S^{5}$ background corresponds to a gauge theory with less
supersymmetry classified in \cite{hep-th/9503121}. This deformation of the
string background can be obtained applying two T-dualities accompanied by
certain shift parametrized by $\beta$ (TsT transformation). For real values of
$\beta$, Frolov obtained the Lax operator for the deformed background which
proves the integrability at classical level \cite{hep-th/0503201}. String
theory in this background was studied in \cite{hep-th/0503192,hep-th/0506063}
and its pp-wave limit was investigated in \cite{hep-th/0505227,hep-th/0505243}%
. The $\beta$-deformations of more complicated (non)supersymmetric backgrounds
was considered also in \cite{hep-th/0505100,hep-th/0507021}.

The aim of this paper is to consider a simple extension of the transformations
considered in \cite{hep-th/0502086,hep-th/0503201} and to prove that under
TsT-transformations applied to any background possessing $U(1)$ symmetries,
the corresponding currents before and after the transformation are equal.

The paper is organized as follows. In the next section we give a brief review
of the $\beta$-deformations of the $\mathcal{N}=4$ gauge theory and its
gravity dual. In section 3 we consider a general background with at least two
$U(1)$ isometries. We show that the $U(1)$ currents are equal after
$Ts_{1}s_{2}\cdots s_{n}T$ transformations where $s_{1}\cdots s_{n}$ means
multi shifts along the remaining $U(1)$ variables. In the next section, as an
example for multi shift procedure, we consider $AdS_{5}\times S^{5}$ and find
a new background parametrized by two real parameters. We show that the new
background reduces to those found in \cite{hep-th/0502086,hep-th/0503201} when
one of the parameters vanishes. We also consider the limit of point-like
string which corresponds actually to BMN limit. In the concluding section we
comment on the results found in the paper.

\section{ Lunin-Maldacena background}

In this Section we give a very brief review of the procedure of Lunin and
Maldacena for obtaining the gravity dual of the $\beta$-deformed SYM theory
considered in \cite{hep-th/9503121}.

Let us consider the $\mathcal{N}=4$ SYM gauge theory in terms of
$\mathcal{N}=1$ SUSY. The theory contains a vector multiplet $V$ and three
chiral multiplets $\Phi^{i}$. The superpotential is given by the expression
\begin{equation}
W=g^{\prime}Tr\left[  [\Phi^{1},\Phi^{2}]\Phi^{3}\right]  . \label{i1}%
\end{equation}
The action then can be written as
\begin{align}
S=  &  Tr\left\lbrace \int d^{4}x d^{4}\theta e^{-gV}\bar\Phi_{i} e^{gV}%
\Phi^{i} + \frac{1}{2g^{2}}\left[  \int d^{4}x d^{2}\theta W^{\alpha}%
W_{\alpha}+c.c.\right]  \right. \nonumber\\
&  \left.  \frac{g^{\prime}}{3!}\left[  \int d^{4} d^{2}\theta\varepsilon
_{ijk}\Phi^{i} [\Phi^{j},\Phi^{k}] + c.c. \right]  \right\rbrace .
\end{align}
We note that the $\mathcal{N}=4$ theory is conformal at any value of the
complex coupling
\begin{equation}
\tau=\frac{\theta}{2\pi}+ \frac{4\pi i}{g^{2}_{YM}}%
\end{equation}
and the deformations that change this value are exactly marginal.

In \cite{hep-th/9503121} Leigh and Strassler considered deformations of the
superpotential of the form
\begin{equation}
W=h\,Tr\left[  e^{i\pi\beta}\Phi_{1}\Phi_{2}\Phi_{3}-e^{-i\pi\beta}\Phi
_{1}\Phi_{3}\Phi_{2}\right]  +h^{\prime}\,Tr\left[  \Phi_{1}^{3}+\Phi_{2}%
^{3}+\Phi_{3}^{3}\right]
\end{equation}

Let us focus on $h^{\prime}=0$ case. The symmetries are: one $U(1)$ R-symmetry
group and two global $U(1)\times U(1)$ groups acting as follows
\begin{align}
&  U(1)_{1}:\quad(\Phi_{1},\Phi_{2}\Phi_{3})\rightarrow(\Phi_{1}%
,e^{i\varphi_{1}}\Phi_{2},e^{-i\varphi_{1}}\Phi_{3})\nonumber\\
&  U(1)_{2}:\quad(\Phi_{1},\Phi_{2}\Phi_{3})\rightarrow(e^{-i\varphi_{2}}%
\Phi_{1},e^{i\varphi_{2}}\Phi_{2},\Phi_{3}).
\end{align}
Since the theory is periodic in $\beta$ one can think of $\beta$ as living on
a torus with complex structure $\tau_{s}$ and the $Sl(2,Z)$ duality group acts
on it and $\beta$ as follows:
\begin{align}
&  \tau_{s}\rightarrow\frac{a\tau_{s}+b}{c\tau_{s}+d}; \quad\beta
\rightarrow\frac{\beta}{c\tau_{s}+d}\nonumber\\
&  \beta\sim\beta+1\sim\beta+\tau_{s}.
\end{align}
As a result of all this we end up with a $\mathcal{N}=1$ SCFT theory.

The gravity dual for real $\beta$ can be obtained in three steps
\cite{hep-th/0503201}. Consider the $S^{5}$ part of $AdS_{5}\times S^{5}$
background. In the first step we perform a T-duality with respect to one of
the $U(1)$ isometries parametrized by the angle $\varphi_{1}$\footnote{See the
Appendix for general $U(1)$ T-duality}. The second step consists in performing
a shift $\varphi_{2}\rightarrow\varphi_{2}+\gamma\varphi_{1}$ where
$\varphi_{2}$ parametrizes another $U(1)$ isometry of the background and
$\gamma$ is a real parameter. In the last step we T-dualize back on
$\varphi_{1}$. The resulting geometry is described by
\begin{equation}
ds_{str}^{2}=R^{2}\left[  ds_{AdS_{5}}^{2}+\sum(dr_{i}^{2}+Gr_{i}^{2}d\phi
_{i}^{2})+\tilde{\gamma}^{2}r_{1}^{2}r_{2}^{2}r_{3}^{2}(\sum d\phi_{i}%
)^{2}\right]  ,
\end{equation}
where
\begin{equation}
G^{-1}=1+\gamma^{2}(r_{1}^{2}r_{2}^{2}+r_{2}^{2}r_{3}^{2}+r_{1}^{2}r_{3}%
^{2});\\
\ \tilde{\gamma}=R^{2}\gamma.
\end{equation}
The other fields are correspondingly\footnote{See for details
\cite{hep-th/0502086,hep-th/0503201}}
\begin{align}
&  e^{2\phi}=e^{2\phi_{0}}G\\
&  B^{NS}=\tilde{\gamma}^{2}R^{2}G(r_{1}^{2}r_{2}^{2}d\phi_{1}d\phi_{2}%
+r_{2}^{2}r_{3}^{2}d\phi_{2}d\phi_{3}+r_{3}^{2}r_{1}^{2}d\phi_{3}d\phi_{1})\\
&  C_{2}=-3\gamma(16\pi N)w_{1}d\psi\\
&  C_{4}=(16\pi N)w_{4}+Gw_{1}d\phi_{1}d\phi2d\phi_{3})\\
&  F_{5}=(16\pi N)(w_{AdS_{5}}+Gw_{S^{5}}).
\end{align}
Using the fact that the currents $J_{\alpha}$ before the TsT-transformations
are equal to the currents $\tilde{J}_{\alpha}$ after the transformations,
Frolov obtained the Lax operator for the deformed geometry, thus proving the
integrability at classical level. The properties of string theory in this
background were further studied in \cite{hep-th/0503192,hep-th/0506063}.
Penrose limit of the Lunin-Maldacena background was investigated in
\cite{hep-th/0505227,hep-th/0505243}.

\section{$U\left(  1\right)  $ Currents and $TsT$ transformation}

As mentioned in the previous section, based on the observation that the string
$U(1)$ currents before and after TsT-transformation are equal, Frolov was able
to obtain the Lax operator of the theory in the deformed background. He also
conjectured that the equality of the currents holds for any two backgrounds
related by TsT-transformation. Below we prove the following

\begin{description}
\item[Proposition:] The $U\left(  1\right)  $ currents of strings in any two
backgrounds related by TsT transformation are equal.
\end{description}

We start with the general action%
\begin{equation}
S=-\frac{\sqrt{\lambda}}{2}\int d\tau\frac{d\sigma}{2\pi}\left[
\gamma^{\alpha\beta}\partial_{\alpha}X^{\mu}\partial_{\beta}X^{\nu}G_{\mu\nu
}-\epsilon^{\alpha\beta}\partial_{\alpha}X^{\mu}\partial_{\beta}X^{\nu}%
B_{\mu\nu}\right]  . \label{1}%
\end{equation}
We will assume that $G_{\mu\nu}$\ and $B_{\mu\nu}$\ do not depend on $X^{1}%
$\ and $X^{2}$\ allowing to perform TsT transformation.

In what follows we use the notations $\mu=1,\cdots,d$, $i=2,\cdots,d$,
$a=3,\cdots,d$. We will prove the statement in several steps.

\noindent Step 1: T-duality on $X^{1}$.

For completeness we write again the T-duality rules and relations\footnote{See
also the Appendix}
\begin{align}
\tilde{G}_{11}  &  =\frac{1}{G_{11}}\text{, \ \ }\tilde{G}_{ij}=G_{ij}%
-\frac{G_{1i}G_{1j}-B_{1i}B_{1j}}{G_{11}}\text{, \ \ }\tilde{G}_{1i}%
=\frac{B_{1i}}{G_{11}},\nonumber\\
\tilde{B}_{ij}  &  =B_{ij}-\frac{G_{1i}B_{1j}-B_{1i}G_{1j}}{G_{11}}\text{,
\ \ }\tilde{B}_{1i}=\frac{G_{1i}}{G_{11}}, \label{2}%
\end{align}%
\begin{align}
\epsilon^{\alpha\beta}\partial_{\beta}\tilde{X}^{1}  &  =\gamma^{\alpha\beta
}\partial_{\beta}X^{M}G_{1M}-\epsilon^{\alpha\beta}\partial_{\beta}X^{M}%
B_{1M},\label{3}\\
\partial_{\alpha}\tilde{X}^{1}  &  =\gamma_{\alpha\sigma}\epsilon^{\sigma\rho
}\partial_{\rho}X^{\mu}G_{1\mu}-\partial_{\alpha}X^{\mu}B_{1\mu},\label{4}\\
\partial_{\alpha}X^{1}  &  =\gamma_{\alpha\sigma}\epsilon^{\sigma\rho}%
\partial_{\rho}\tilde{X}^{\mu}\tilde{G}_{1\mu}-\partial_{\alpha}\tilde{X}%
^{\mu}\tilde{B}_{1\mu},\label{5}\\
\tilde{X}^{i}  &  =X^{i}. \label{6}%
\end{align}
The T-dual action has the same form but with transformed background fields%
\begin{equation}
\tilde{S}=-\frac{\sqrt{\lambda}}{2}\int d\tau\frac{d\sigma}{2\pi}\left[
\gamma^{\alpha\beta}\partial_{\alpha}\tilde{X}^{\mu}\partial_{\beta}\tilde
{X}^{\nu}\tilde{G}_{\mu\nu}-\epsilon^{\alpha\beta}\partial_{\alpha}\tilde
{X}^{\mu}\partial_{\beta}\tilde{X}^{\nu}\tilde{B}_{\mu\nu}\right]  .
\end{equation}
\noindent Step 2 consists in shift of $\tilde{X}^{2}$%
\begin{align}
\tilde{X}^{2}  &  =\tilde{x}^{2}+\hat{\gamma}\tilde{x}^{1},\nonumber\\
\tilde{X}^{1}  &  =\tilde{x}^{1}\text{, \ \ }\tilde{X}^{a}=\tilde{x}^{a}.
\end{align}
Note that the background remains independent of $\tilde{X}^{1}$ and $\tilde
{X}^{2}$.

The shift described above produces the following transformations of the metric%
\begin{align}
\tilde{g}_{11}  &  =\tilde{G}_{11}+2\hat{\gamma}\tilde{G}_{12}+\hat{\gamma
}^{2}\tilde{G}_{22},\nonumber\\
\tilde{g}_{1i}  &  =\tilde{G}_{1i}+\hat{\gamma}\tilde{G}_{2i},\nonumber\\
\tilde{g}_{ij}  &  =\tilde{G}_{ij},
\end{align}
and for the $\tilde{B}_{\mu\nu}$ we get%
\begin{align}
\tilde{b}_{ij}  &  =\tilde{B}_{ij},\nonumber\\
\tilde{b}_{1i}  &  \rightarrow\tilde{B}_{1i}+\hat{\gamma}\tilde{B}_{2i}.
\end{align}
The relations (\ref{3}-\ref{5}) are also changed, for instance (\ref{5})
becomes%
\begin{equation}
\partial_{\alpha}X^{1}=\gamma_{\alpha\sigma}\epsilon^{\sigma\rho}%
\partial_{\rho}\tilde{x}^{\mu}\tilde{G}_{1\mu}-\partial_{\alpha}\tilde{x}%
^{\mu}\tilde{B}_{1\mu}+\hat{\gamma}\gamma_{\alpha\sigma}\epsilon^{\sigma\rho
}\partial_{\rho}\tilde{x}^{1}\tilde{G}_{12}-\hat{\gamma}\partial_{\alpha
}\tilde{x}^{1}\tilde{B}_{12}. \label{relation1}%
\end{equation}
Note that it is crucial that the background is independent of $X^{1}$\ and
$X^{2}$, otherwise we cannot perform a T-duality back on $\tilde{x}_{1}$.

In the new variables the action is given by%
\begin{equation}
\bar{\tilde{S}}=-\frac{\sqrt{\lambda}}{2}\int d\tau\frac{d\sigma}{2\pi}\left[
\gamma^{\alpha\beta}\partial_{\alpha}\tilde{x}^{\mu}\partial_{\beta}\tilde
{x}^{\nu}\tilde{g}_{\mu\nu}-\epsilon^{\alpha\beta}\partial_{\alpha}\tilde
{x}^{\mu}\partial_{\beta}\tilde{x}^{\nu}\tilde{b}_{\mu\nu}\right]  .
\end{equation}

\noindent In step 3 we T-dualize back on $\tilde{x}^{1}$.

The action again has the standard form%
\begin{equation}
\tilde{\bar{\tilde{S}}}=-\frac{\sqrt{\lambda}}{2}\int d\tau\frac{d\sigma}%
{2\pi}\left[  \gamma^{\alpha\beta}\partial_{\alpha}x^{\mu}\partial_{\beta
}x^{\nu}g_{\mu\nu}-\epsilon^{\alpha\beta}\partial_{\alpha}x^{\mu}%
\partial_{\beta}x^{\nu}b_{\mu\nu}\right]  , \label{12}%
\end{equation}
where $g_{\mu\nu}$\ and $b_{\mu\nu}$\ are obtained from $\tilde{g}_{\mu\nu}%
$\ and $\tilde{b}_{\mu\nu}$ \ by making use of the standard rules eqs.
(\ref{2}-\ref{5}).

Now we will prove that the currents $J_{\mu}^{\alpha}$ and $j_{\mu}^{\alpha}%
$\ obtained from (\ref{1}) and (\ref{12}) respectively are equal, i.e.%
\begin{equation}
J_{\mu}^{\alpha}=j_{\mu}^{\alpha}, \label{equality}%
\end{equation}
where%
\begin{align}
j_{\mu}^{\alpha}  &  =-\sqrt{\lambda}\gamma^{\alpha\beta}\partial_{\beta
}x^{\nu}g_{\mu\nu}+\sqrt{\lambda}\epsilon^{\alpha\beta}\partial_{\beta}x^{\nu
}b_{\mu\nu},\\
J_{\mu}^{\alpha}  &  =-\sqrt{\lambda}\gamma^{\alpha\beta}\partial_{\beta
}x^{\nu}G_{\mu\nu}+\sqrt{\lambda}\epsilon^{\alpha\beta}\partial_{\beta}x^{\nu
}B_{\mu\nu}.
\end{align}
We will prove the statement directly, but in two steps.

\noindent a) First we will prove the equality (\ref{equality}) for
$J_{1}^{\alpha}$ and $j_{1}^{\alpha}$\ and then for $J_{i}^{\alpha}$ and
$j_{i}^{\alpha}$%
\begin{align}
\frac{j_{1}^{\alpha}}{-\sqrt{\lambda}}  &  =\gamma^{\alpha\beta}%
\partial_{\beta}x^{1}g_{11}+\gamma^{\alpha\beta}\partial_{\beta}x^{i}%
g_{1i}-\epsilon^{\alpha\beta}\partial_{\beta}x^{i}b_{1i}\nonumber\\
&  =\gamma^{\alpha\beta}\partial_{\beta}x^{1}g_{11}+\gamma^{\alpha\beta
}\partial_{\beta}\tilde{x}^{i}g_{1i}-\epsilon^{\alpha\beta}\partial_{\beta
}\tilde{x}^{i}b_{1i}\nonumber\\
&  =\frac{\gamma^{\alpha\beta}}{\tilde{g}_{11}}\left(  \gamma_{\beta\sigma
}\epsilon^{\sigma\rho}\partial_{\rho}\tilde{x}^{\mu}\tilde{g}_{1\mu}%
-\partial_{\beta}\tilde{x}^{\mu}\tilde{b}_{1\mu}\right)  +\gamma^{\alpha\beta
}\partial_{\beta}\tilde{x}^{i}\frac{\tilde{b}_{1i}}{\tilde{g}_{11}}%
-\epsilon^{\alpha\beta}\partial_{\beta}\tilde{x}^{i}\frac{\tilde{g}_{1i}%
}{\tilde{g}_{11}}\nonumber\\
&  =\gamma^{\alpha\beta}\gamma_{\beta\sigma}\epsilon^{\sigma\rho}%
\partial_{\rho}\tilde{x}^{\mu}\frac{\tilde{g}_{1\mu}}{\tilde{g}_{11}}%
-\epsilon^{\alpha\beta}\partial_{\beta}\tilde{x}^{i}\frac{\tilde{g}_{1i}%
}{\tilde{g}_{11}}\nonumber\\
&  =\epsilon^{\alpha\beta}\partial_{\beta}\tilde{x}^{1}.
\end{align}
Now we use (\ref{3}) and find%
\begin{equation}
\frac{j_{1}^{\alpha}}{-\sqrt{\lambda}}=\gamma^{\alpha\beta}\partial_{\beta
}X^{\mu}G_{1\mu}-\epsilon^{\alpha\beta}\partial_{\beta}X^{\mu}B_{1\mu}%
=\frac{J_{1}^{\alpha}}{-\sqrt{\lambda}}.
\end{equation}
\noindent b) We turn now to the case of $J_{i}^{\alpha}$ and $j_{i}^{\alpha}$
$\left(  i=2,\cdots,d\right)  $. In this case there are more transformations
to be performed but all of them are based on (\ref{2}-\ref{5})%
\begin{align}
\frac{j_{i}^{\alpha}}{-\sqrt{\lambda}}  &  =\gamma^{\alpha\beta}%
\partial_{\beta}x^{\mu}g_{i\mu}-\epsilon^{\alpha\beta}\partial_{\beta}x^{\mu
}b_{i\mu}\nonumber\\
&  =\gamma^{\alpha\beta}\partial_{\beta}x^{1}g_{i1}+\gamma^{\alpha\beta
}\partial_{\beta}\tilde{x}^{j}g_{ij}-\epsilon^{\alpha\beta}\partial_{\beta
}x^{1}b_{1i}-\epsilon^{\alpha\beta}\partial_{\beta}\tilde{x}^{j}%
b_{ij}\nonumber\\
&  =\gamma^{\alpha\beta}\partial_{\beta}\tilde{x}^{1}\tilde{g}_{i1}%
+\gamma^{\alpha\beta}\partial_{\beta}\tilde{x}^{j}\tilde{g}_{ij}%
+\epsilon^{\alpha\beta}\partial_{\beta}\tilde{x}^{1}\tilde{b}_{1i}%
-\epsilon^{\alpha\beta}\partial_{\beta}\tilde{x}^{j}\tilde{b}_{ij}.
\end{align}
Now we go to the $\tilde{X}^{\mu}$ variables by making the inverse shift%
\begin{equation}
\frac{j_{i}^{\alpha}}{-\sqrt{\lambda}}=\gamma^{\alpha\beta}\partial_{\beta
}\tilde{X}^{\mu}\tilde{G}_{i\mu}-\epsilon^{\alpha\beta}\partial_{\beta}%
\tilde{X}^{\mu}\tilde{B}_{i\mu}.
\end{equation}
Since $\tilde{X}^{i}=X^{i}$, we separate $\tilde{X}^{1}$\ and $\tilde{X}^{i}%
$\ dependent parts and find%
\begin{align}
\frac{j_{i}^{\alpha}}{-\sqrt{\lambda}}  &  =\gamma^{\alpha\beta}%
\partial_{\beta}\tilde{X}^{1}\tilde{G}_{i1}-\epsilon^{\alpha\beta}%
\partial_{\beta}\tilde{X}^{1}\tilde{B}_{i1}+\gamma^{\alpha\beta}%
\partial_{\beta}\tilde{X}^{j}\tilde{G}_{ij}-\epsilon^{\alpha\beta}%
\partial_{\beta}\tilde{X}^{j}\tilde{B}_{ij}\\
&  =.\gamma^{\alpha\beta}\partial_{\beta}X^{1}G_{i1}-\epsilon^{\alpha\beta
}\partial_{\beta}X^{1}B_{i1}+\gamma^{\alpha\beta}\partial_{\beta}X^{j}%
G_{ij}-\epsilon^{\alpha\beta}\partial_{\beta}X^{j}B_{ij}.
\end{align}
Therefore%
\begin{equation}
\frac{j_{i}^{\alpha}}{-\sqrt{\lambda}}=.\gamma^{\alpha\beta}\partial_{\beta
}X^{\mu}G_{i\mu}-\epsilon^{\alpha\beta}\partial_{\beta}X^{\mu}B_{i\mu}%
=\frac{J_{i}^{\alpha}}{-\sqrt{\lambda}},
\end{equation}
which proves the statement (\ref{equality}).%

%TCIMACRO{\TeXButton{equation number}{\setcounter{equation}{0}
%\renewcommand{\theequation}{\arabic{section}.\arabic{equation}}}}%
%BeginExpansion
\setcounter{equation}{0}
\renewcommand{\theequation}{\arabic{section}.\arabic{equation}}%
%EndExpansion

\section{$Ts_{1}\cdots s_{d}T$ transformations}

In this section we make a simple generalization of the TsT-transformation. We
proceed as follows. First we make a T-duality on $X^{1}$ after which the
original action%
\begin{equation}
S=-\frac{\sqrt{\lambda}}{2}\int d\tau\frac{d\sigma}{2\pi}\left[
\gamma^{\alpha\beta}\partial_{\alpha}X^{\mu}\partial_{\beta}X^{\nu}G_{\mu\nu
}-\epsilon^{\alpha\beta}\partial_{\alpha}X^{\mu}\partial_{\beta}X^{\nu}%
B_{\mu\nu}\right]
\end{equation}
becomes%
\begin{equation}
S=-\frac{\sqrt{\lambda}}{2}\int d\tau\frac{d\sigma}{2\pi}\left[
\gamma^{\alpha\beta}\partial_{\alpha}\tilde{X}^{\mu}\partial_{\beta}\tilde
{X}^{\nu}\tilde{G}_{\mu\nu}-\epsilon^{\alpha\beta}\partial_{\alpha}\tilde
{X}^{\mu}\partial_{\beta}\tilde{X}^{\nu}\tilde{B}_{\mu\nu}\right]  ,
\end{equation}
where the tilde variables are defined in (\ref{2}), with the relations
(\ref{4}) and (\ref{6}) satisfied.

Second step consists in applying multi-shifts along the $U(1)$ isometries
unaffected by the T-duality in the previous step. This slightly generalizes
the Maldacena-Lunin procedure described in the previous section,%
\begin{align}
\tilde{X}^{i}  &  =\tilde{x}^{i}+\gamma^{i}\tilde{x}^{1},\nonumber\\
\tilde{X}^{1}  &  =\tilde{x}^{1}, \label{relation2}%
\end{align}
or $\boldsymbol{\tilde{X}}=\boldsymbol{A}\boldsymbol{\tilde{x}}$ where%
\begin{equation}
\boldsymbol{\tilde{X}}=\left(
\begin{array}
[c]{c}%
\tilde{X}^{1}\\
\vdots\\
\tilde{X}^{N}%
\end{array}
\right)  ,\text{ \ \ }\boldsymbol{A}=\left(
\begin{array}
[c]{ccccc}%
1 & 0 & \cdots &  & 0\\
\gamma^{2} & 1 &  &  & \vdots\\
\vdots &  & \ddots &  & 0\\
\gamma^{N} & 0 & \cdots & 0 & 1
\end{array}
\right)  .
\end{equation}
Under these multi-shifts the background fields take the form%
\begin{align}
\tilde{g}_{11}  &  =\tilde{G}_{11}+2\gamma^{i}\tilde{G}_{1i}+\gamma^{i}%
\gamma^{j}\tilde{G}_{ij},\nonumber\\
\tilde{g}_{1i}  &  =\tilde{G}_{1i}+\gamma^{j}\tilde{G}_{ij},\nonumber\\
\tilde{g}_{ij}  &  =\tilde{G}_{ij},\nonumber\\
\tilde{b}_{1i}  &  =\tilde{B}_{1i}+\gamma^{j}\tilde{B}_{ij},\nonumber\\
\tilde{b}_{ij}  &  =\tilde{B}_{ij}.
\end{align}
The last step consists in T-dualization back on $\tilde{x}^{1}$. The resulting
action is%
\begin{equation}
S=-\frac{\sqrt{\lambda}}{2}\int d\tau\frac{d\sigma}{2\pi}\left[
\gamma^{\alpha\beta}\partial_{\alpha}x^{\mu}\partial_{\beta}x^{\nu}g_{\mu\nu
}-\epsilon^{\alpha\beta}\partial_{\alpha}x^{\mu}\partial_{\beta}x^{\nu}%
b_{\mu\nu}\right]  .
\end{equation}

As in the case of TsT-transformation, for the generalization described above
we prove below

\begin{description}
\item[Proposition:] The $U\left(  1\right)  $ currents of strings in any two
backgrounds related by $Ts_{1}\cdots S_{n} T$ transformation are equal.
\end{description}

\noindent Proof: One can first consider $j_{1}^{\alpha}$ and using the
relations between the variables write them in terms of the original
coordinates%
\begin{align}
\frac{j_{1}^{\alpha}}{-\sqrt{\lambda}}  &  =\gamma^{\alpha\beta}%
\partial_{\beta}x^{\gamma}g_{1\gamma}-\epsilon^{\alpha\beta}\partial_{\beta
}x^{i}b_{1i}\nonumber\\
&  =\gamma^{\alpha\beta}\partial_{\beta}x^{1}g_{11}+\gamma^{\alpha\beta
}\partial_{\beta}\tilde{x}^{i}g_{1i}-\epsilon^{\alpha\beta}\partial_{\beta
}\tilde{x}^{i}b_{1i}\nonumber\\
&  =\frac{\gamma^{\alpha\beta}}{\tilde{g}_{11}}\left(  \gamma_{\beta\sigma
}\epsilon^{\sigma\rho}\partial_{\rho}\tilde{x}^{\mu}\tilde{g}_{1\mu}%
-\partial_{\beta}\tilde{x}^{\mu}\tilde{b}_{1\mu}\right)  +\gamma^{\alpha\beta
}\partial_{\beta}\tilde{x}^{i}\frac{\tilde{b}_{1i}}{\tilde{g}_{11}}%
-\epsilon^{\alpha\beta}\partial_{\beta}\tilde{x}^{i}\frac{\tilde{g}_{1i}%
}{\tilde{g}_{11}}\nonumber\\
&  =\gamma^{\alpha\beta}\gamma_{\beta\sigma}\epsilon^{\sigma\rho}%
\partial_{\rho}\tilde{x}^{\mu}\frac{\tilde{g}_{1\mu}}{\tilde{g}_{11}}%
-\epsilon^{\alpha\beta}\partial_{\beta}\tilde{x}^{i}\frac{\tilde{g}_{1i}%
}{\tilde{g}_{11}}\nonumber\\
&  =\epsilon^{\alpha\beta}\partial_{\beta}\tilde{x}^{1}.
\end{align}
But%
\begin{equation}
\frac{J_{1}^{\alpha}}{-\sqrt{\lambda}}=\gamma^{\alpha\beta}\partial_{\beta
}X^{\mu}G_{1\mu}-\epsilon^{\alpha\beta}\partial_{\beta}X^{\mu}B_{1\mu
}=\epsilon^{\alpha\beta}\partial_{\beta}\tilde{x}^{1},
\end{equation}
and therefore
\begin{equation}
j_{1}^{\alpha}=J_{1}^{\alpha}.
\end{equation}
Let us show that this equality is satisfied for the other currents. One can
easily show that%
\begin{equation}
j_{i}^{\alpha}=\tilde{j}_{i}^{\alpha}.
\end{equation}
Let us see how $\tilde{j}_{i}^{\alpha}$\ is related to $\tilde{J}_{i}^{\alpha
}$%
\begin{align}
\frac{j_{i}^{\alpha}}{-\sqrt{\lambda}}  &  =\gamma^{\alpha\beta}%
\partial_{\beta}\tilde{x}^{\mu}\tilde{g}_{i\mu}-\epsilon^{\alpha\beta}%
\partial_{\beta}\tilde{x}^{\mu}\tilde{b}_{i\mu}\nonumber\\
&  =\gamma^{\alpha\beta}\partial_{\beta}\tilde{x}^{1}\left(  \tilde{G}%
_{1i}+\gamma^{j}\tilde{G}_{ij}\right)  -\gamma^{\alpha\beta}\left(
\partial_{\beta}\tilde{x}^{j}-\gamma^{j}\partial_{\beta}\tilde{x}^{1}\right)
\tilde{G}_{ij}\\
&  +\epsilon^{\alpha\beta}\partial_{\beta}\tilde{x}^{1}\left(  \tilde{B}%
_{1i}+\gamma^{j}\tilde{B}_{ij}\right)  -\epsilon^{\alpha\beta}\left(
\partial_{\beta}\tilde{x}^{j}-\gamma^{j}\partial_{\beta}\tilde{x}^{1}\right)
\tilde{B}_{ij}\\
&  =\gamma^{\alpha\beta}\partial_{\beta}\tilde{x}^{\mu}\tilde{G}_{1\mu
}-\epsilon^{\alpha\beta}\partial_{\beta}\tilde{x}^{\mu}\tilde{B}_{1\mu
}\nonumber\\
&  =\frac{\tilde{J}_{i}^{\alpha}}{-\sqrt{\lambda}}.
\end{align}
Simple calculations now lead to $J_{i}^{\alpha}=\tilde{J}_{i}^{\alpha}$. This
proves that%
\begin{equation}
j_{i}^{\alpha}=J_{i}^{\alpha}.
\end{equation}
Although the proof is straightforward, it may have important consequences. For
instance, if the theory in the initial background is integrable, one can study
integrability of the second theory by making use of the above relation. We
will comment on this issue in the next section.

The equality between the currents in the $AdS_{5}\times S^{5}$ background and
its deformation relate the boundary conditions imposed on the fields in the
initial and the transformed backgrounds. It remains to examine how the
boundary conditions for $x^{\mu}$ and $X^{\mu}$\ in our case are related.
First we notice that the time component of $J_{\mu}^{\alpha}$, i.e. $J_{\mu
}^{0}$ is just the momentum conjugated to $X^{\mu}$. The equality of $j_{\mu
}^{0}$\ and $J_{\mu}^{0}$\ means that the two momenta are equal and constant
(due to the isometry). Therefore this property, observed first in
\cite{hep-th/0503201}, continues to hold in the general case of TsT- and
multi-shift transformations. To examine the boundary conditions we will use
the relation%
\begin{equation}
\partial_{\alpha}x^{1}=\gamma_{\alpha\beta}\epsilon^{\beta\gamma}%
\partial_{\gamma}\tilde{x}^{\mu}\tilde{g}_{1\mu}-\partial_{\alpha}\tilde
{x}^{\mu}\tilde{b}_{1\mu}. \label{tansfrom}%
\end{equation}
To simplify the calculation we choose the conformal gauge for the 2d metric
$\gamma_{\alpha\beta}=diag\left(  -1,1\right)  $ and $\epsilon^{01}=1$. Let us
compute the boundary conditions for $x^{1}$. To do this we need expressions
for the metric components $\tilde{g}_{\mu\nu}$\ in terms of the original
metric $G_{\mu\nu}$. Using the transformation properties we find%
\begin{align}
\tilde{g}_{11}  &  =\frac{G}{G_{11}},\\
\tilde{g}_{1i}  &  =\frac{B_{1i}+\gamma^{j}\left(  G_{ij}G_{11}-G_{1i}%
G_{1j}+B_{1i}B_{1j}\right)  }{G_{11}},
\end{align}
and%
\begin{equation}
\tilde{b}_{1i}=\frac{G_{1i}+\gamma^{j}\left(  B_{ij}G_{11}-G_{1i}B_{1j}%
+B_{1i}G_{1j}\right)  }{G_{11}},
\end{equation}
where%
\begin{equation}
G=1+2\gamma^{i}B_{1i}+\gamma^{i}\gamma^{j}\left(  G_{ij}G_{11}-G_{1i}%
G_{1j}+B_{1i}B_{1j}\right)  ,
\end{equation}
(all others are not changed by the shifts and are given in the Appendix).

Substituting the above expressions for $\tilde{g}_{\mu\nu}$\ and $\tilde
{b}_{\mu\nu}$\ in (\ref{tansfrom}) and using the inverse transformations
relating $\tilde{x}^{\mu}$\ with$X^{\mu}$ , we find%
\begin{equation}
\partial_{1}x^{1}=\partial_{1}X^{1}-\gamma^{i}J_{i}^{0}\text{, \ \ }%
i=2,\cdots,N.
\end{equation}
The boundary conditions for the other coordinates are easily obtained from%
\begin{equation}
\partial_{\alpha}x^{i}=\partial_{\alpha}\tilde{x}^{i}-\gamma^{i}%
\partial_{\alpha}\tilde{x}^{1}.
\end{equation}
Using the relation (\ref{relation1}) and (\ref{relation2}) we get%
\begin{equation}
\partial_{1}x^{i}=\partial_{1}X^{i}+\gamma^{i}\left(  \partial_{0}x^{\mu
}G_{1\mu}+\partial_{1}x^{j}B_{1j}\right)  =\partial_{1}x^{i}+\gamma^{i}%
J_{1}^{0}.
\end{equation}
Therefore, the boundary conditions for the fields in the deformed background
are twisted as follows%
\begin{align}
\partial_{1}x^{1}  &  =\partial_{1}X^{1}-\gamma^{i}J_{i}^{0},
\label{twisted 1}\\
\partial_{1}x^{i}  &  =\partial_{1}X^{i}+\gamma^{i}J_{1}^{0}.
\label{twisted 2}%
\end{align}
Integrating over $\sigma$ we find%
\begin{align}
x^{1}\left(  2\pi\right)  -x^{1}\left(  0\right)   &  =2\pi\left(
n_{1}-\gamma^{i}J_{i}\right)  ,\\
x^{i}\left(  2\pi\right)  -x^{i}\left(  0\right)   &  =2\pi\left(
n_{i}+\gamma^{i}J_{1}\right)  ,
\end{align}
where%
\begin{equation}
X^{\mu}\left(  2\pi\right)  -X^{\mu}\left(  0\right)  =2\pi n_{\mu},
\end{equation}
and the current%
\begin{equation}
J_{\mu}=\int\frac{d\sigma}{2\pi}J_{\mu}^{0}.
\end{equation}
In the next section we will apply these results to the $AdS_{5}\times S^{5}$
background and analyse the implications of these transformations to string theory.%

%TCIMACRO{\TeXButton{equation number}{\setcounter{equation}{0}
%\renewcommand{\theequation}{\arabic{section}.\arabic{equation}}}}%
%BeginExpansion
\setcounter{equation}{0}
\renewcommand{\theequation}{\arabic{section}.\arabic{equation}}%
%EndExpansion

\section{$\left(  \hat{\gamma}_{2},\hat{\gamma}_{3}\right)  $-deformation}

\subsection{Supergravity solution}

We start with the $S^{5}$ part of string action as in \cite{hep-th/0503201}
with $i=1,2,3.$%
\begin{equation}
S=-\frac{\sqrt{\lambda}}{2}\int d\tau\frac{d\sigma}{2\pi}\left[
\gamma^{\alpha\beta}\left(  \partial_{\alpha}r_{i}\partial_{\beta}r_{i}%
+g_{ij}\partial_{\alpha}\tilde{\tilde{\varphi}}_{i}\partial_{\beta}%
\tilde{\tilde{\varphi}}_{j}\right)  +\Lambda\left(  r_{i}^{2}-1\right)
\right]  , \label{S5}%
\end{equation}
where the metric $g_{ij}$ and the anti-symmetric 2-form field $b_{ij}$\ are%
\begin{align}
g_{11}  &  =r_{2}^{2}+r_{3}^{2}\text{, \ \ }g_{22}=r_{1}^{2}+r_{2}^{2}\text{,
\ \ }g_{33}=1,\nonumber\\
g_{12}  &  =r_{2}^{2}\text{, \ \ \ \ \ \ \ \ }g_{13}=r_{2}^{2}-r_{3}%
^{2}\text{, \ \ }g_{23}=r_{2}^{2}-r_{1}^{2},\nonumber\\
b_{ij}  &  =0.
\end{align}
and $\Lambda$ is a Lagrangian multiplier which ensures the constraint%
\begin{equation}
\sum r_{i}^{2}=1. \label{constrain}%
\end{equation}
This action is related to the one used in \cite{hep-th/0502086} by the
following change of the variables
\begin{equation}
\tilde{\tilde{\varphi}}_{1}=\dfrac{1}{3}\left(  \hat{\varphi}_{1}+\hat
{\varphi}_{2}-2\hat{\varphi}_{3}\right)  \text{, \ \ }\tilde{\tilde{\varphi}%
}_{2}=\dfrac{1}{3}\left(  -2\hat{\varphi}_{1}+\hat{\varphi}_{2}+\hat{\varphi
}_{3}\right)  \text{, \ \ }\tilde{\tilde{\varphi}}_{3}=\dfrac{1}{3}\left(
\hat{\varphi}_{1}+\hat{\varphi}_{2}+\hat{\varphi}_{3}\right)  ,
\label{variables}%
\end{equation}
which leads to the following relations between the old and new angular momenta%
\begin{align}
\tilde{\tilde{J}}_{1}  &  =\hat{J}_{2}-\hat{J}_{3},\label{J1}\\
\tilde{\tilde{J}}_{2}  &  =\hat{J}_{2}-\hat{J}_{1},\label{J2}\\
\tilde{\tilde{J}}_{3}  &  =\hat{J}_{1}+\hat{J}_{2}+\hat{J}_{3}. \label{J3}%
\end{align}
We next make the T-duality transformation on the circle parametrized by
$\varphi_{1}$, the action becomes%
\begin{equation}
S=-\frac{\sqrt{\lambda}}{2}\int d\tau\frac{d\sigma}{2\pi}\left[
\gamma^{\alpha\beta}\left(  \partial_{\alpha}r_{i}\partial_{\beta}r_{i}%
+\tilde{g}_{ij}\partial_{\alpha}\tilde{\varphi}_{i}\partial_{\beta}%
\tilde{\varphi}_{j}\right)  -\epsilon^{\alpha\beta}\tilde{b}_{ij}%
\partial_{\alpha}\tilde{\varphi}_{i}\partial_{\beta}\tilde{\varphi}%
_{j}+\Lambda\left(  r_{i}^{2}-1\right)  \right]  ,
\end{equation}
where%
\begin{align}
\tilde{g}_{11}  &  =\frac{1}{r_{2}^{2}+r_{3}^{2}}\text{, \ \ }\tilde{g}%
_{22}=\frac{r_{1}^{2}r_{2}^{2}+r_{1}^{2}r_{3}^{2}+r_{2}^{2}r_{3}^{2}}%
{r_{2}^{2}+r_{3}^{2}}\text{, \ \ }\tilde{g}_{33}=-\frac{r_{2}^{2}+r_{3}%
^{2}-\left(  r_{2}^{2}-r_{3}^{2}\right)  ^{2}}{r_{2}^{2}+r_{3}^{2}%
},\nonumber\\
\tilde{g}_{12}  &  =\tilde{g}_{13}=0\text{, \ \ }\tilde{g}_{23}=\frac
{2r_{2}^{2}r_{3}^{2}-r_{1}^{2}r_{2}^{2}-r_{1}^{2}r_{3}^{2}}{r_{2}^{2}%
+r_{3}^{2}},\nonumber\\
\tilde{b}_{12}  &  =\frac{r_{2}^{2}}{r_{2}^{2}+r_{3}^{2}}\text{, \ \ }%
\tilde{b}_{13}=\frac{r_{2}^{2}-r_{3}^{2}}{r_{2}^{2}+r_{3}^{2}}\text{,
\ \ }\tilde{b}_{23}=0.
\end{align}
The T-dual variables $\tilde{\varphi}_{i}$ are related to $\tilde
{\tilde{\varphi}}_{i}$ as follows%
\begin{align}
\partial_{\alpha}\tilde{\tilde{\varphi}}_{1}  &  =\gamma_{\alpha\beta}%
\epsilon^{\beta\gamma}\partial_{\gamma}\tilde{\varphi}_{1}\tilde{g}%
_{11}-\partial_{\alpha}\tilde{\varphi}_{i}\tilde{b}_{1i},\nonumber\\
\tilde{\tilde{\varphi}}_{2}  &  =\tilde{\varphi}_{2}\text{, \ \ }\tilde
{\tilde{\varphi}}_{3}=\tilde{\varphi}_{3}. \label{tansfrom 1}%
\end{align}
Next, we make the following shift of the angle variables $\tilde{\varphi}_{2}$
and $\tilde{\varphi}_{3}$ simultaneously%
\begin{equation}
\tilde{\varphi}_{2}\rightarrow\tilde{\varphi}_{2}+\hat{\gamma}_{2}%
\tilde{\varphi}_{1},\text{ \ \ }\tilde{\varphi}_{3}\rightarrow\tilde{\varphi
}_{3}+\hat{\gamma}_{3}\tilde{\varphi}_{1},
\end{equation}
where $\hat{\gamma}_{2}$ and $\hat{\gamma}_{3}$ are two arbitrary constants.
The metric transforms in the following way under the above shift%
\begin{align}
\tilde{g}_{11}  &  \rightarrow\tilde{g}_{11}+\hat{\gamma}_{2}^{2}\tilde
{g}_{22}+\hat{\gamma}_{3}^{2}\tilde{g}_{33}+2\hat{\gamma}_{2}\tilde{g}%
_{12}+2\hat{\gamma}_{3}\tilde{g}_{13}+2\hat{\gamma}_{2}\hat{\gamma}_{3}%
\tilde{g}_{23},\nonumber\\
\tilde{g}_{12}  &  \rightarrow\tilde{g}_{12}+\hat{\gamma}_{2}\tilde{g}%
_{22}+\hat{\gamma}_{3}\tilde{g}_{23},\nonumber\\
\tilde{g}_{13}  &  \rightarrow\tilde{g}_{13}+\hat{\gamma}_{2}\tilde{g}%
_{23}+\hat{\gamma}_{3}\tilde{g}_{33},\nonumber\\
\tilde{b}_{12}  &  \rightarrow\tilde{b}_{12}-\hat{\gamma}_{3}\tilde{b}%
_{23},\nonumber\\
\tilde{b}_{13}  &  \rightarrow\tilde{b}_{13}+\hat{\gamma}_{2}\tilde{b}_{23},
\end{align}
and the variables $\tilde{\varphi}_{i}$ transforms into%
\begin{align}
\partial_{\alpha}\tilde{\tilde{\varphi}}_{1}  &  =\gamma_{\alpha\beta}%
\epsilon^{\beta\gamma}\partial_{\gamma}\tilde{\varphi}_{1}\tilde{g}%
_{11}-\partial_{\alpha}\tilde{\varphi}_{i}\tilde{b}_{1i}-\hat{\gamma}%
_{2}\partial_{\alpha}\tilde{\varphi}_{1}\tilde{b}_{12}-\hat{\gamma}%
_{3}\partial_{\alpha}\tilde{\varphi}_{1}\tilde{b}_{13},\nonumber\\
\tilde{\tilde{\varphi}}_{2}  &  =\tilde{\varphi}_{2}\text{, \ \ }\tilde
{\tilde{\varphi}}_{3}=\tilde{\varphi}_{3}. \label{tansfrom 2}%
\end{align}
Finally, we make the T-duality transformation on the circle parametrized by
$\tilde{\varphi}_{1}$ again. After the TsT transformation, the $\left(
\hat{\gamma}_{2},\hat{\gamma}_{3}\right)  $-deformed background becomes%
\begin{equation}
S=-\frac{\sqrt{\lambda}}{2}\int d\tau\frac{d\sigma}{2\pi}\left[
\gamma^{\alpha\beta}\left(  \partial_{\alpha}r_{i}\partial_{\beta}r_{i}%
+G_{ij}\partial_{\alpha}\varphi_{i}\partial_{\beta}\varphi_{i}\right)
-\epsilon^{\alpha\beta}B_{ij}\partial_{\alpha}\varphi_{i}\partial_{\beta
}\varphi_{i}+\Lambda\left(  r_{i}^{2}-1\right)  \right]  , \label{sigma}%
\end{equation}
where%

\begin{align}
G_{1i}  &  =Gg_{1i},\nonumber\\
G_{22}  &  =G\left(  g_{22}+9\hat{\gamma}_{3}^{2}r_{1}^{2}r_{2}^{2}r_{3}%
^{2}\right)  ,\nonumber\\
G_{33}  &  =G\left(  g_{33}+9\hat{\gamma}_{2}^{2}r_{1}^{2}r_{2}^{2}r_{3}%
^{2}\right)  ,\nonumber\\
G_{23}  &  =G\left(  g_{23}-9\hat{\gamma}_{2}\hat{\gamma}_{3}r_{1}^{2}%
r_{2}^{2}r_{3}^{2}\right)  ,\nonumber\\
B_{12}  &  =G\left[  \hat{\gamma}_{2}\left(  r_{1}^{2}r_{2}^{2}+r_{1}^{2}%
r_{3}^{2}+r_{2}^{2}r_{3}^{2}\right)  +\hat{\gamma}_{3}\left(  2r_{2}^{2}%
r_{3}^{2}-r_{1}^{2}r_{2}^{2}-r_{1}^{2}r_{3}^{2}\right)  \right]  ,\nonumber\\
B_{13}  &  =G\left[  \hat{\gamma}_{2}\left(  2r_{2}^{2}r_{3}^{2}-r_{1}%
^{2}r_{2}^{2}-r_{1}^{2}r_{3}^{2}\right)  +\hat{\gamma}_{3}\left(  r_{2}%
^{2}+r_{3}^{2}-\left(  r_{2}^{2}-r_{3}^{2}\right)  ^{2}\right)  \right]
,\nonumber\\
B_{23}  &  =-G\left[  \hat{\gamma}_{2}\left(  2r_{1}^{2}r_{2}^{2}-r_{1}%
^{2}r_{3}^{2}-r_{2}^{2}r_{3}^{2}\right)  +\hat{\gamma}_{3}\left(  g_{13}%
g_{23}-g_{12}\right)  \right]  , \label{background}%
\end{align}
where%
\begin{equation}
G^{-1}=1+\hat{\gamma}_{2}^{2}\left(  r_{1}^{2}r_{2}^{2}+r_{1}^{2}r_{3}%
^{2}+r_{2}^{2}r_{3}^{2}\right)  +\hat{\gamma}_{3}^{2}\left[  r_{2}^{2}%
+r_{3}^{2}-\left(  r_{2}^{2}-r_{3}^{2}\right)  ^{2}\right]  +2\hat{\gamma}%
_{2}\hat{\gamma}_{3}\left(  2r_{2}^{2}r_{3}^{2}-r_{1}^{2}r_{2}^{2}-r_{1}%
^{2}r_{3}^{2}\right)  ,
\end{equation}
and we have used the constraint (\ref{constrain}).

The variables $\tilde{\varphi}_{i}$ are related to the T-dual variables
$\varphi_{i}$ as follows%
\begin{align}
\partial_{\alpha}\tilde{\varphi}_{1}  &  =\gamma_{\alpha\beta}\epsilon
^{\beta\gamma}\partial_{\gamma}\tilde{\varphi}_{i}G_{1i}-\partial_{\alpha
}\tilde{\varphi}_{i}B_{1i},\nonumber\\
\tilde{\varphi}_{2}  &  =\varphi_{2}\text{, \ \ }\tilde{\varphi}_{3}%
=\varphi_{3}. \label{tansfrom 3}%
\end{align}
The equations (\ref{tansfrom 1}), (\ref{tansfrom 2}) and (\ref{tansfrom 3})
allow us to determine the following relations between the angle variables
$\tilde{\tilde{\varphi}}_{i}$\ and the TsT-transformed variables $\varphi_{i}%
$:%
\begin{align}
\partial_{\alpha}\tilde{\tilde{\varphi}}_{1}  &  =\left[  \tilde{g}_{11}%
G_{1i}+\left(  \hat{\gamma}_{2}\tilde{b}_{12}+\hat{\gamma}_{3}\tilde{b}%
_{13}\right)  B_{1i}-\tilde{b}_{1i}\right]  \partial_{\alpha}\varphi
_{i}\nonumber\\
&  -\left[  \left(  \hat{\gamma}_{2}\tilde{b}_{12}+\hat{\gamma}_{3}\tilde
{b}_{13}\right)  G_{1i}+\tilde{g}_{11}B_{1i}\right]  \gamma_{\alpha\beta
}\epsilon^{\beta\gamma}\partial_{\gamma}\varphi_{i},\\
\partial_{\alpha}\tilde{\tilde{\varphi}}_{2}  &  =\partial_{\alpha}\varphi
_{2}-\hat{\gamma}_{2}B_{1i}\partial_{\alpha}\varphi_{i}+\hat{\gamma}_{2}%
G_{1i}\gamma_{\alpha\beta}\epsilon^{\beta\gamma}\partial_{\gamma}\varphi
_{i},\\
\partial_{\alpha}\tilde{\tilde{\varphi}}_{3}  &  =\partial_{\alpha}\varphi
_{3}-\hat{\gamma}_{3}B_{1i}\partial_{\alpha}\varphi_{i}+\hat{\gamma}_{3}%
G_{1i}\gamma_{\alpha\beta}\epsilon^{\beta\gamma}\partial_{\gamma}\varphi_{i},
\end{align}
which gives the boundary conditions%
\begin{align}
\tilde{\tilde{\varphi}}_{1}^{\prime}  &  =\varphi_{1}^{\prime}+\hat{\gamma
}_{2}J_{2}^{0}+\hat{\gamma}_{3}J_{3}^{0},\nonumber\\
\tilde{\tilde{\varphi}}_{2}^{\prime}  &  =\varphi_{2}^{\prime}-\hat{\gamma
}_{2}J_{1}^{0},\nonumber\\
\tilde{\tilde{\varphi}}_{3}^{\prime}  &  =\varphi_{3}^{\prime}-\hat{\gamma
}_{3}J_{1}^{0}, \label{boundary}%
\end{align}
which are consistent with the boundary conditions (\ref{twisted 1}) and
(\ref{twisted 2}). It is easy to see that when $\hat{\gamma}_{3}=0$, the above
background reduces to the Lunin-Maldacena background
\cite{hep-th/0502086,hep-th/0503201}.

We can check that the Virasoro constraint%
\begin{equation}
g_{ij}\left(  \dot{\tilde{\tilde{\varphi}}}_{i}\dot{\tilde{\tilde{\varphi}}%
}_{j}+\tilde{\tilde{\varphi}}_{i}^{\prime}\tilde{\tilde{\varphi}}_{j}^{\prime
}\right)  =G_{ij}\left(  \dot{\varphi}_{i}\dot{\varphi}_{j}+\varphi
_{i}^{\prime}\varphi_{j}^{\prime}\right)  ,
\end{equation}
is satisfied as expected.

\subsection{The dual field theory}

According to the AdS/CFT duality, string theory in the background
(\ref{background}) is dual a field theory on the boundary of the AdS space.
This field theory is a deformed theory from $\mathcal{N}$=4 SYM theory by the
deformation $\left(  \hat{\gamma}_{2},\hat{\gamma}_{3}\right)  $, so we will
call it $\left(  \hat{\gamma}_{2},\hat{\gamma}_{3}\right)  $-deformed
$\mathcal{N}$=4 SYM theory. Now the question is: what is this dual field
theory? To answer this question, let us look at the symmetries of the deformed
background (\ref{background}).

We try first to find how many supersymmetries are preserved in the dual field
theory. To derive the background (\ref{background}), we wrote the $S^{5}$ part
of $AdS\times S^{5}$ as (\ref{S5}). The metric has manifestly a $U\left(
1\right)  \times U\left(  1\right)  \times U\left(  1\right)  $ isometry, of
which a $U\left(  1\right)  \times U\left(  1\right)  $ preserve the Killing
spinors. In the case of Lunin-Maldacena background, a very special torus was
chosen to compactify the 10d string theory. The $TsT$ transformation only
breaks the supersymmetry corresponding to the Killing spinor associated to
$U\left(  1\right)  \times U\left(  1\right)  $ so that the deformed
background preserves 1/4 supersymmetries. The left $U\left(  1\right)  $
remains an R-symmetry in the dual $\mathcal{N}$ = 1 SYM theory. In our case,
$TssT$ transformation breaks all $U\left(  1\right)  \times U\left(  1\right)
\times U\left(  1\right)  $ isometry so that no Killing spinor is preserved.
Therefore the dual field theory has \emph{no} supersymmetry!

Next we try to learn more about the dual field theory from the gravity side.
Let us recall the relation between the TsT transformation of the supergravity
background and the star product of the dual field theory in the case of
Lunin-Maldacena background \cite{hep-th/0502086}. $SL\left(  2,R\right)  $
acts on the parameter
\begin{equation}
\tau=B_{12}+i\sqrt{g},
\end{equation}
as%
\begin{equation}
\tau\rightarrow\tau^{\prime}=\frac{\tau}{1+\gamma\tau}\text{ \ or \ }\frac
{1}{\tau}\rightarrow\frac{1}{\tau^{\prime}}=\frac{1}{\tau}+\gamma.
\label{SL(2,R)}%
\end{equation}
Schematically, $1/\tau$ can be written as \cite{hep-th/9908142}%
\begin{equation}
\frac{1}{\tau}\sim\left(  \frac{1}{\boldsymbol{g}+\boldsymbol{B}}\right)
^{ij}=G_{\text{open}}^{ij}+\theta^{ij},
\end{equation}
where $G_{\text{open}}^{ij}$ is the open string metric and $\theta^{ij}$\ is
the noncommutative parameter. Then the result of the $SL\left(  2,R\right)
$\ transformation (\ref{SL(2,R)}) is just to introduce a noncommutatity
parameter $\theta^{12}\sim\gamma$. This analogy can be seen more precisely if
we define a $2\times2$ matrix $\boldsymbol{\gamma}$ as%
\begin{equation}
\boldsymbol{\gamma}\equiv\left(  \frac{1}{\boldsymbol{g}^{\prime
}+\boldsymbol{B}^{\prime}}\right)  -\left(  \frac{1}{\boldsymbol{g}%
+\boldsymbol{B}}\right)  =\left(  \boldsymbol{G}_{\text{open}}^{\prime
}-\boldsymbol{G}_{\text{open}}\right)  +\left(  \boldsymbol{\theta}^{\prime
}-\boldsymbol{\theta}\right)  . \label{Gamma}%
\end{equation}
It is easy to get the matrix
\begin{equation}
\boldsymbol{\gamma}=\left(
\begin{array}
[c]{cc}%
0 & -\gamma\\
\gamma & 0
\end{array}
\right)  .
\end{equation}
Thus the TsT transformation of the supergravity background is equivalent to a
shift of the noncommutative parameter by $\theta^{12}=-\gamma$ in the dual
field theory.

Now let us look at the $\left(  \hat{\gamma}_{2},\hat{\gamma}_{3}\right)
$-deformed background which we found in the previous section. We can similarly
define a $3\times3$ matrix $\boldsymbol{\gamma}$ as in (\ref{Gamma}).
Straightforward calculation leads to the following\footnote{Here we define new
symbols $\left(  \gamma^{12},\gamma^{13}\right)  $\ which are related to the
symbols we used in the previous section as $\gamma^{12}\equiv\hat{\gamma}%
^{2}/R^{2}$ and $\gamma^{13}\equiv\hat{\gamma}^{3}/R^{2}$, where $R$\ is the
radius of $S^{5}$.} $\boldsymbol{\gamma}$%
\begin{equation}
\boldsymbol{\gamma}=\left(
\begin{array}
[c]{ccc}%
0 & -\gamma^{12} & -\gamma^{13}\\
\gamma^{12} & 0 & 0\\
\gamma^{13} & 0 & 0
\end{array}
\right)  .
\end{equation}
Thus in our case the TsT transformation of the supergravity background is
equivalent to a shift of the noncommutative parameters by $\theta^{12}%
=-\gamma^{12}$ and $\theta^{13}=-\gamma^{13}$ in the dual field theory. Since
the modification only affects the directions $\left(  \phi_{1},\phi_{2}%
,\phi_{3}\right)  $, the action of the dual field theory will be the same as
the one of the $\mathcal{N}$=4 SYM theory except the superpotential term,
which can be obtained from the undeformed one by replacing the usual product
$\phi_{i}\phi_{j}$ by the associative star product $\phi_{i}\ast\phi_{j}$.
Obviously, we will not be able to write down the action by using the
$\mathcal{N}$=1 superfields since all supersymmetries are broken in the process.

\subsection{Semiclassical analysis}

A classical solution of the sigma model associated with the background
(\ref{background}) is obtained as%
\begin{align}
t  &  =\tau\text{, \ \ }\rho=0,\nonumber\\
\varphi_{1}  &  =\nu_{1}\tau\text{, \ \ }\varphi_{2}=\nu_{2}\tau\text{,
\ \ }\varphi_{3}=\nu_{3}\tau,\nonumber\\
\alpha &  =\arccos\left(  \sqrt{\frac{\hat{\gamma}_{2}+2\hat{\gamma}_{3}%
}{4\hat{\gamma}_{3}-\hat{\gamma}_{2}}}\right)  \text{, \ \ }\theta=\frac{\pi
}{4}, \label{classic solution}%
\end{align}
where%
\begin{equation}
\nu_{1}=-\frac{2}{3}\text{, \ \ }\nu_{2}=\frac{4}{3}\text{, \ \ }\nu_{3}%
=\frac{1}{3}.
\end{equation}
The angular momenta and the energy corresponding to this state are%
\begin{align}
J_{1}  &  =0,\\
J_{2}  &  =-\dfrac{3\hat{\gamma}_{3}}{\hat{\gamma}_{2}-4\hat{\gamma}_{3}}C,\\
J_{3}  &  =\dfrac{3\hat{\gamma}_{2}}{\hat{\gamma}_{2}-4\hat{\gamma}_{3}}C,
\end{align}
and%
\begin{equation}
E=\nu_{1}J_{1}+\nu_{2}J_{2}+\nu_{3}J_{3}=3C,
\end{equation}
where $C\left(  \propto N\right)  $\ is a constant. From the relations of
angular momenta (\ref{J1}-\ref{J3}), we can see that this solution is
associated to the state with
\begin{equation}
\left(  \hat{J}_{1},\hat{J}_{2},\hat{J}_{3}\right)  =\left(  \dfrac
{\hat{\gamma}_{2}+2\hat{\gamma}_{3}}{\hat{\gamma}_{2}-4\hat{\gamma}_{3}%
}C,\dfrac{\hat{\gamma}_{2}-\hat{\gamma}_{3}}{\hat{\gamma}_{2}-4\hat{\gamma
}_{3}}C,\dfrac{\hat{\gamma}_{2}-\hat{\gamma}_{3}}{\hat{\gamma}_{2}%
-4\hat{\gamma}_{3}}C\right)  .
\end{equation}
It is easy to see that the above state reduces to $\left(  J,J,J\right)
$\ state when $\hat{\gamma}_{3}=0$\ and to $\left(  -J,0,0\right)  $\ state
when $\hat{\gamma}_{2}=\hat{\gamma}_{3}$ with $J=E/3$.

Next, let us consider the fluctuations around the above classical solution
(\ref{classic solution}) with large 't Hooft coupling $\lambda=g_{\text{YM}%
}N=R^{4}/\alpha^{\prime2}$ as%
\begin{align}
t  &  =\tau+\frac{1}{\lambda^{1/4}}\tilde{t}\text{, \ \ }\rho=\frac{1}%
{\lambda^{1/4}}\tilde{\rho},\nonumber\\
\varphi_{1}  &  =\nu_{1}\tau+\frac{1}{\lambda^{1/4}}\tilde{\varphi}_{1}\text{,
\ \ }\varphi_{2}=\nu_{2}\tau+\frac{1}{\lambda^{1/4}}\tilde{\varphi}_{2}\text{,
\ \ }\varphi_{3}=\nu_{3}\tau+\frac{1}{\lambda^{1/4}}\tilde{\varphi}%
_{3},\nonumber\\
\alpha &  =\arccos\left(  \sqrt{\frac{\hat{\gamma}_{2}+2\hat{\gamma}_{3}%
}{4\hat{\gamma}_{3}-\hat{\gamma}_{2}}}\right)  +\frac{1}{\lambda^{1/4}}%
\tilde{\alpha}\text{, \ \ }\theta=\frac{\pi}{4}+\frac{1}{\lambda^{1/4}}%
\sqrt{\frac{\hat{\gamma}_{2}-4\hat{\gamma}_{3}}{2\left(  \hat{\gamma}_{2}%
-\hat{\gamma}_{3}\right)  }}\tilde{\theta}.
\end{align}
where we have defined%
\begin{align*}
r_{1}  &  =\cos\alpha,\\
r_{2}  &  =\sin\alpha\cos\theta,\\
r_{3}  &  =\sin\alpha\sin\theta.
\end{align*}
The difference between energy and angular momenta is%
\begin{equation}
E-\left(  \nu_{1}J_{1}+\nu_{2}J_{2}+\nu_{3}J_{3}\right)  =\frac{1}{2}\int
_{0}^{2\pi}\frac{d\sigma}{2\pi}\mathcal{H}%
\end{equation}
where the energy and angular momenta are defined as%
\begin{align}
E  &  \equiv P_{t}=-\frac{\delta S}{\delta\dot{t}}\text{,}\\
J_{i}  &  \equiv P_{\varphi_{i}}=\frac{\delta S}{\delta\dot{\varphi}_{i}%
},\text{ \ \ }i=1,2,3,
\end{align}
and $\mathcal{H}$ is the corresponding Hamiltonian. By using the Virasoro
constraints%
\begin{equation}
T_{aa}=G_{mn}\partial_{a}X^{m}\partial_{a}X^{n}=0,
\end{equation}
and keeping the terms up to quadratic order, the transverse Hamiltonian can be
obtained as%

\begin{align}
\mathcal{H}  &  =-\partial_{a}\tilde{t}\partial_{a}\tilde{t}+\eta_{\mu}%
^{2}+\partial_{a}\eta_{\mu}\partial_{a}\eta_{\mu}\nonumber\\
&  +4G\left(  \hat{\gamma}_{2}+2\hat{\gamma}_{3}\right)  \left(  \hat{\gamma
}_{2}-\hat{\gamma}_{3}\right)  ^{2}\tilde{\alpha}^{2}+\partial_{a}%
\tilde{\alpha}\partial_{a}\tilde{\alpha}+4G\left(  \hat{\gamma}_{2}%
+2\hat{\gamma}_{3}\right)  \left(  \hat{\gamma}_{2}-\hat{\gamma}_{3}\right)
^{2}\tilde{\theta}^{2}+\partial_{a}\tilde{\theta}\partial_{a}\tilde{\theta
}\nonumber\\
&  +2G\left(  \hat{\gamma}_{3}-\hat{\gamma}_{2}\right)  \partial_{a}%
\tilde{\varphi}_{1}\partial_{a}\tilde{\varphi}_{1}\nonumber\\
&  +\frac{3G\hat{\gamma}_{3}}{\left(  4\hat{\gamma}_{3}-\hat{\gamma}%
_{2}\right)  ^{2}}\left(  3\hat{\gamma}_{2}^{3}\hat{\gamma}_{3}+\hat{\gamma
}_{2}^{2}-9\hat{\gamma}_{2}\hat{\gamma}_{3}^{3}+6\hat{\gamma}_{3}^{4}%
-8\hat{\gamma}_{2}\hat{\gamma}_{3}+16\hat{\gamma}_{3}^{2}\right)  \partial
_{a}\tilde{\varphi}_{2}\partial_{a}\tilde{\varphi}_{2}\nonumber\\
&  +\frac{G}{\left(  4\hat{\gamma}_{3}-\hat{\gamma}_{2}\right)  ^{2}}\left(
9\hat{\gamma}_{2}^{5}+64\hat{\gamma}_{3}^{3}+18\hat{\gamma}_{2}^{2}\hat
{\gamma}_{3}^{3}+12\hat{\gamma}_{2}^{2}\hat{\gamma}_{3}-\hat{\gamma}_{2}%
^{3}-27\hat{\gamma}_{2}^{3}\hat{\gamma}_{3}^{2}-48\hat{\gamma}_{2}\hat{\gamma
}_{3}^{2}\right)  \partial_{a}\tilde{\varphi}_{3}\partial_{a}\tilde{\varphi
}_{3}\nonumber\\
&  +2G\left(  \hat{\gamma}_{3}-\hat{\gamma}_{2}\right)  \partial_{a}%
\tilde{\varphi}_{1}\partial_{a}\tilde{\varphi}_{2}\nonumber\\
&  +\dfrac{2G}{\left(  4\hat{\gamma}_{3}-\hat{\gamma}_{2}\right)  ^{2}}\left(
15\hat{\gamma}_{2}^{2}\hat{\gamma}_{3}-18\hat{\gamma}_{2}\hat{\gamma}_{3}%
^{4}-16\hat{\gamma}_{3}^{3}+27\hat{\gamma}_{2}^{2}\hat{\gamma}_{3}^{3}%
-2\hat{\gamma}_{2}^{3}-9\hat{\gamma}_{2}^{4}\hat{\gamma}_{3}-24\hat{\gamma
}_{2}\hat{\gamma}_{3}^{2}\right)  \partial_{a}\tilde{\varphi}_{2}\partial
_{a}\tilde{\varphi}_{3}\nonumber\\
&  +2G\left(  \hat{\gamma}_{3}-\hat{\gamma}_{2}\right)  \sqrt{2\left(
\hat{\gamma}_{3}-\hat{\gamma}_{2}\right)  \left(  \hat{\gamma}_{2}%
+2\hat{\gamma}_{3}\right)  }\tilde{\alpha}\left(  \tilde{\varphi}_{2}^{\prime
}+2\tilde{\varphi}_{1}^{\prime}\right) \nonumber\\
&  +2G\left(  \hat{\gamma}_{3}-\hat{\gamma}_{2}\right)  \left(  \hat{\gamma
}_{2}+2\hat{\gamma}_{3}\right)  \sqrt{\dfrac{2\left(  \hat{\gamma}_{3}%
-\hat{\gamma}_{2}\right)  }{4\hat{\gamma}_{3}-\hat{\gamma}_{2}}}\tilde{\theta
}\left(  \tilde{\varphi}_{2}^{\prime}-4\tilde{\varphi}_{3}^{\prime}\right)  .
\end{align}
where we have made a change of coordinates $\left(  \tilde{\rho},\Omega
_{3}\right)  \rightarrow\eta_{\mu}$, $\mu=1,2,3,4$ and%
\begin{equation}
G^{-1}=\hat{\gamma}_{2}^{3}-\hat{\gamma}_{2}-3\hat{\gamma}_{2}\hat{\gamma}%
_{3}^{2}+2\hat{\gamma}_{3}^{2}+4\hat{\gamma}_{3}.
\end{equation}
We diagonalize the Hamiltonian by making the following coordinates
transformations,%
\begin{align*}
\tilde{\varphi}_{1}  &  =\phi_{1}-\frac{1}{2}\phi_{2},\\
\tilde{\varphi}_{2}  &  =\phi_{2},\\
\tilde{\varphi}_{3}  &  =\phi_{3}-\frac{15\hat{\gamma}_{2}^{2}\hat{\gamma}%
_{3}-18\hat{\gamma}_{2}\hat{\gamma}_{3}^{4}-16\hat{\gamma}_{3}^{3}%
+27\hat{\gamma}_{2}^{2}\hat{\gamma}_{3}^{3}-2\hat{\gamma}_{2}^{3}-9\hat
{\gamma}_{2}^{4}\hat{\gamma}_{3}-24\hat{\gamma}_{2}\hat{\gamma}_{3}^{2}}%
{9\hat{\gamma}_{2}^{5}+64\hat{\gamma}_{3}^{3}+18\hat{\gamma}_{2}^{2}%
\hat{\gamma}_{3}^{3}+12\hat{\gamma}_{2}^{2}\hat{\gamma}_{3}-\hat{\gamma}%
_{2}^{3}-27\hat{\gamma}_{2}^{3}\hat{\gamma}_{3}^{2}-48\hat{\gamma}_{2}%
\hat{\gamma}_{3}^{2}}\phi_{2}.
\end{align*}
Then,%
\begin{align}
\mathcal{H}  &  =-\partial_{a}\tilde{t}\partial_{a}\tilde{t}+\eta_{\mu}%
^{2}+\partial_{a}\eta_{\mu}\partial_{a}\eta_{\mu}\nonumber\\
&  +4G\left(  \hat{\gamma}_{2}+2\hat{\gamma}_{3}\right)  \left(  \hat{\gamma
}_{2}-\hat{\gamma}_{3}\right)  ^{2}\tilde{\alpha}^{2}+\partial_{a}%
\tilde{\alpha}\partial_{a}\tilde{\alpha}+4G\left(  \hat{\gamma}_{2}%
+2\hat{\gamma}_{3}\right)  \left(  \hat{\gamma}_{2}-\hat{\gamma}_{3}\right)
^{2}\tilde{\theta}^{2}+\partial_{a}\tilde{\theta}\partial_{a}\tilde{\theta
}\nonumber\\
&  +2G\left(  \hat{\gamma}_{3}-\hat{\gamma}_{2}\right)  \partial_{a}\phi
_{1}\partial_{a}\phi_{1}\nonumber\\
&  +\frac{9G^{2}\left(  \hat{\gamma}_{3}-\hat{\gamma}_{2}\right)  \left(
4\hat{\gamma}_{3}-\hat{\gamma}_{2}\right)  }{9\hat{\gamma}_{2}^{5}%
+64\hat{\gamma}_{3}^{3}+18\hat{\gamma}_{2}^{2}\hat{\gamma}_{3}^{3}%
+12\hat{\gamma}_{2}^{2}\hat{\gamma}_{3}-\hat{\gamma}_{2}^{3}-27\hat{\gamma
}_{2}^{3}\hat{\gamma}_{3}^{2}-48\hat{\gamma}_{2}\hat{\gamma}_{3}^{2}}%
\partial_{a}\phi_{2}\partial_{a}\phi_{2}\nonumber\\
&  +\frac{2G}{\left(  4\hat{\gamma}_{3}-\hat{\gamma}_{2}\right)  }\left(
9\hat{\gamma}_{2}^{5}+64\hat{\gamma}_{3}^{3}+18\hat{\gamma}_{2}^{2}\hat
{\gamma}_{3}^{3}+12\hat{\gamma}_{2}^{2}\hat{\gamma}_{3}-\hat{\gamma}_{2}%
^{3}-27\hat{\gamma}_{2}^{3}\hat{\gamma}_{3}^{2}-48\hat{\gamma}_{2}\hat{\gamma
}_{3}^{2}\right)  \partial_{a}\phi_{3}\partial_{a}\phi_{3}\nonumber\\
&  +4G\left(  \hat{\gamma}_{2}-\hat{\gamma}_{3}\right)  \left(  \hat{\gamma
}_{2}-4\hat{\gamma}_{3}\right)  \sqrt{2\left(  \hat{\gamma}_{2}-\hat{\gamma
}_{3}\right)  \left(  \hat{\gamma}_{2}+2\hat{\gamma}_{3}\right)  }%
\tilde{\alpha}\phi_{1}^{\prime}\nonumber\\
&  -2G\left(  \hat{\gamma}_{2}-\hat{\gamma}_{3}\right)  \left(  \hat{\gamma
}_{2}+2\hat{\gamma}_{3}\right)  \sqrt{2\left(  \hat{\gamma}_{2}-\hat{\gamma
}_{3}\right)  \left(  \hat{\gamma}_{2}-4\hat{\gamma}_{3}\right)  }%
\tilde{\theta}\nonumber\\
&  \cdot\left(  \frac{9\hat{\gamma}_{2}\left(  4\hat{\gamma}_{3}-\hat{\gamma
}_{2}\right)  }{9\hat{\gamma}_{2}^{5}+64\hat{\gamma}_{3}^{3}+18\hat{\gamma
}_{2}^{2}\hat{\gamma}_{3}^{3}+12\hat{\gamma}_{2}^{2}\hat{\gamma}_{3}%
-\hat{\gamma}_{2}^{3}-27\hat{\gamma}_{2}^{3}\hat{\gamma}_{3}^{2}-48\hat
{\gamma}_{2}\hat{\gamma}_{3}^{2}}\phi_{2}^{\prime}+4\phi_{3}^{\prime}\right)
.
\end{align}
Since the coefficients are constants, the Hamiltonian can be quantized to get
the string spectrum as discussed in \cite{hep-th/0304198,hep-th/0507021}

\section{Conclusions}

In this paper, we consider a deformation of the $AdS_{5}\times S^{5}$
background of string theory. We propose a simple generalization of the
Lunin-Maldacena procedure for obtaining a so called beta deformed theory
which, from the gauge theory side, corresponds to a deformation of Yang-Mills
theory studied by Leigh and Strassler. For real deformation parameter
$\beta=\gamma$, the Lunin-Maldacena background can be thought of as a
T-duality on one of the angles $\phi_{1}$ corresponding to one of the three
$U(1)$ isometries of $AdS_{5}\times S^{5}$ background , a shift on another
isometry variable, followed by T-duality again of $\phi_{1}$. It was proved in
the original paper by Lunin and Maldacena that this procedure does not produce
additional singularities except for only those in the original background. Our
generalization consists in additional shifts on the other $U(1)$ variables in
the intermediate step. In this way one can obtain a new deformed background
which depends on more parameters $\gamma_{1}\cdots\gamma_{n}$. Since this
procedure consists only in additional shifts, the resulting background again
contains only the singularities descended from the original one.

In Section II, we reviewed the Lunin-Maldacena background and the
TsT-transformation procedure. In the next Section we have proved that the
currents for any two backgrounds related by TsT-transformations are equal
(which was conjectured in \cite{hep-th/0503201}).

In the next Section, we consider Ts...sT-transformations. We find that due to
these transformations the boundary conditions for the $U(1)$ variables are
twisted. We prove also that the $U(1)$ currents in any two backgrounds related
by Ts...sT-transformations are equal. This property is important since, as it
is discussed in \cite{hep-th/0503201}, it means that the theory preserves the
nice property of integrability. The integrability can be proved along the
lines of the paper by Frolov \cite{hep-th/0503201}.

In Section V, we apply the TssT-transformation to $AdS_{5}\times S^{5}$
background. The obtained background is new and the string theory on it is
integrable. We argue that the supersymmetry is broken and the background is
less supersymmetric than that of Lunin and Maldacena.

After short comments on the gauge theory side, we perform a semiclassical
analysis of string theory in $\gamma_{2}-\gamma_{3}$ deformed $AdS_{5}\times
S^{5}$ background. We study the theory in the BMN limit and obtain the
corresponding conserved quantities important for AdS/CFT correspondence. It is
important to note that for $\gamma_{3}=0$ the background and therefore string
theory, reduce to that studied by Lunin and Maldacena. In the Appendix we give
for completeness a detailed derivation of the T-duality transformations.

There are several ways to develop the results obtained in this paper. First of
all one can study multi spin solutions in our background along the lines of
\cite{hep-th/0503192}. To clarify the AdS/CFT correspondence one must consider
the gauge theory side in more detail. It would be interesting to see what kind
of spin chain should describe the string and gauge theory in this case. One
can use then the powerful Bethe ansatz technique to study the correspondence
on both sides. We leave these questions for further study.

\begin{acknowledgments}
We would like thank J. Maldacena and C. N\'{u}\~{n}ez for reading our draft
and giving many useful suggestions. We also thank S. Frolov for pointing out a
mistake in our first version. R.R and Y.Y thank Simon Fraser University for
kind hospitality. This work is supported by an operating grant from the
Natural Sciences and Engineering Research Council of Canada. Y.Y. is also
partially supported by a postdoctoral fellowship from the National Science
Council of Taiwan.
\end{acknowledgments}

\appendix%

%TCIMACRO{\TeXButton{equation number in appendix}{\setcounter{equation}{0}
%\renewcommand{\theequation}{\thesection.\arabic{equation}}}}%
%BeginExpansion
\setcounter{equation}{0}
\renewcommand{\theequation}{\thesection.\arabic{equation}}%
%EndExpansion

\section{T-duality transformations}

In this Appendix we give detailed derivation of the T-duality transformation.

We start with the general string theory action:%
\begin{equation}
S=-\frac{\sqrt{\lambda}}{2}\int d\tau\frac{d\sigma}{2\pi}\left[
\gamma^{\alpha\beta}\partial_{\alpha}X^{M}\partial_{\beta}X^{N}G_{MN}\left(
X^{i}\right)  -\epsilon^{\alpha\beta}\partial_{\alpha}X^{M}\partial_{\beta
}X^{N}B_{MN}\left(  X^{i}\right)  \right]  , \label{action}%
\end{equation}
where: a) $M,N =1,\cdots,d-1$, $i=2,\cdots,d-1$ and b) the background fields
$G_{MN}$\ and $B_{MN}$\ do not depend on $X^{1}$.

The equation of motion for $X^{1}$ tells us that there exists conserved
current $J^{\alpha}$:%
\begin{equation}
\partial_{\alpha}J^{\alpha}=0\Leftrightarrow J^{\alpha}\equiv-\frac
{\sqrt{\lambda}}{2\pi}\frac{\partial\mathcal{L}}{\partial\left(
\partial_{\alpha}X^{1}\right)  }.
\end{equation}
Let us define $p^{\alpha}$ as:%
\begin{equation}
p^{\alpha}=\gamma^{\alpha\beta}\partial_{\beta}X^{N}G_{1N}-\epsilon
^{\alpha\beta}\partial_{\beta}X^{N}B_{1N}. \label{p}%
\end{equation}
The action (\ref{constrain}) can be rewritten in terms of $p^{\alpha}$ as
follows:%
\begin{align}
S  &  =-\sqrt{\lambda}\int d\tau\frac{d\sigma}{2\pi}\left[  \frac{1}{2}%
\gamma^{\alpha\beta}\partial_{\alpha}X^{1}\partial_{\beta}X^{1}G_{11}%
+\gamma^{\alpha\beta}\partial_{\alpha}X^{1}\partial_{\beta}X^{i}%
G_{1i}-\epsilon^{\alpha\beta}\partial_{\alpha}X^{1}\partial_{\beta}X^{N}%
B_{1N}\right. \nonumber\\
&  \left.  +\frac{1}{2}\left(  \gamma^{\alpha\beta}\partial_{\alpha}%
X^{i}\partial_{\beta}X^{j}G_{ij}-\epsilon^{\alpha\beta}\partial_{\alpha}%
X^{i}\partial_{\beta}X^{j}B_{ij}\right)  \right] \nonumber\\
&  =-\sqrt{\lambda}\int d\tau\frac{d\sigma}{2\pi}\left[  \partial_{\alpha
}X^{1}\left(  \gamma^{\alpha\beta}\partial_{\beta}X^{N}G_{1N}-\epsilon
^{\alpha\beta}\partial_{\beta}X^{N}B_{1N}\right)  -\frac{1}{2}\gamma
^{\alpha\beta}\partial_{\alpha}X^{1}\partial_{\beta}X^{1}G_{11}\right.
\nonumber\\
&  \left.  +\frac{1}{2}\left(  \gamma^{\alpha\beta}\partial_{\alpha}%
X^{i}\partial_{\beta}X^{j}G_{ij}-\epsilon^{\alpha\beta}\partial_{\alpha}%
X^{i}\partial_{\beta}X^{j}B_{ij}\right)  \right] \nonumber\\
&  =-\sqrt{\lambda}\int d\tau\frac{d\sigma}{2\pi}\left[  p^{\alpha}%
\partial_{\alpha}X^{1}-\frac{1}{2}\gamma^{\alpha\beta}\partial_{\alpha}%
X^{1}\partial_{\beta}X^{1}G_{11}+\frac{1}{2}\left(  \gamma^{\alpha\beta
}\partial_{\alpha}X^{i}\partial_{\beta}X^{j}G_{ij}-\epsilon^{\alpha\beta
}\partial_{\alpha}X^{i}\partial_{\beta}X^{j}B_{ij}\right)  \right]  .
\label{action-1}%
\end{align}
Let us consider the second term in the above expression%
\begin{equation}
\frac{1}{2}\gamma^{\alpha\beta}\partial_{\alpha}X^{1}\partial_{\beta}%
X^{1}G_{11}=\partial_{\alpha}X^{1}G_{11}\frac{\gamma^{\alpha\beta}}{2G_{11}%
}\partial_{\beta}X^{1}G_{11}. \label{2nd term}%
\end{equation}
In order to perform T-duality, we have to eliminate $X^{1}$ which enters the
action only through $\partial_{\alpha}X^{1}G_{11}$. From the definition of
$p^{\alpha}$:%
\begin{equation}
p^{\alpha}=\gamma^{\alpha\beta}\partial_{\beta}X^{1}G_{11}+\gamma^{\alpha
\beta}\partial_{\beta}X^{i}G_{1i}-\epsilon^{\alpha\beta}\partial_{\beta}%
X^{i}B_{1i},
\end{equation}
we find:%
\begin{equation}
\gamma^{\alpha\beta}\partial_{\beta}X^{1}G_{11}=p^{\alpha}-\gamma^{\alpha
\beta}\partial_{\beta}X^{i}G_{1i}+\epsilon^{\alpha\beta}\partial_{\beta}%
X^{i}B_{1i}.
\end{equation}
Substituting for $\partial_{\alpha}X^{1}G_{11}$ in (\ref{2nd term}) we find:%
\begin{align}
&  \partial_{\alpha}X^{1}G_{11}\frac{\gamma^{\alpha\beta}}{2G_{11}}%
\partial_{\beta}X^{1}G_{11}\nonumber\\
&  =\gamma^{\alpha\sigma}\partial_{\sigma}X^{1}G_{11}\frac{\gamma_{\alpha
\beta}}{2G_{11}}\gamma^{\beta\rho}\partial_{\rho}X^{1}G_{11}\nonumber\\
&  =\left(  p^{\alpha}-\gamma^{\alpha\sigma}\partial_{\sigma}X^{i}%
G_{1i}+\epsilon^{\alpha\sigma}\partial_{\sigma}X^{i}B_{1i}\right)
\frac{\gamma_{\alpha\beta}}{2G_{11}}\left(  p^{\beta}-\gamma^{\beta\rho
}\partial_{\rho}X^{i}G_{1i}+\epsilon^{\beta\rho}\partial_{\rho}X^{i}%
B_{1i}\right) \nonumber\\
&  =\frac{p^{\alpha}\gamma_{\alpha\beta}p^{\beta}}{2G_{11}}-p^{\alpha}\left(
\partial_{\alpha}X^{i}\frac{G_{1i}}{G_{11}}-\gamma_{\alpha\beta}%
\epsilon^{\beta\rho}\partial_{\rho}X^{i}\frac{B_{1i}}{G_{11}}\right)
\nonumber\\
&  +\frac{1}{2}\gamma^{\alpha\beta}\partial_{\alpha}X^{i}\partial_{\beta}%
X^{j}\frac{G_{1i}G_{1j}}{G_{11}}+\frac{1}{2}\epsilon^{\alpha\sigma}%
\gamma_{\alpha\beta}\epsilon^{\beta\rho}\partial_{\sigma}X^{i}\partial_{\rho
}X^{j}\frac{B_{1i}B_{1j}}{G_{11}}\nonumber\\
&  +\frac{\epsilon^{\alpha\sigma}}{2}\gamma_{\alpha\beta}\gamma^{\beta\rho
}\partial_{\rho}X^{j}\partial_{\sigma}X^{i}\frac{B_{1i}G_{1j}}{G_{11}}%
+\frac{\epsilon^{\beta\rho}}{2}\gamma_{\alpha\beta}\gamma^{\alpha\sigma
}\partial_{\sigma}X^{i}\partial_{\rho}X^{j}\frac{G_{1i}B_{1j}}{G_{11}%
}\nonumber\\
&  =\frac{p^{\alpha}\gamma_{\alpha\beta}p^{\beta}}{2G_{11}}-p^{\alpha}\left(
\partial_{\alpha}X^{i}\frac{G_{1i}}{G_{11}}-\gamma_{\alpha\beta}%
\epsilon^{\beta\rho}\partial_{\rho}X^{i}\frac{B_{1i}}{G_{11}}\right)
\nonumber\\
&  +\frac{1}{2}\left(  \gamma^{\alpha\beta}\partial_{\alpha}X^{i}%
\partial_{\beta}X^{j}\frac{G_{1i}G_{1j}-B_{1i}B_{1j}}{G_{11}}-\epsilon
^{\alpha\beta}\partial_{\alpha}X^{i}\partial_{\beta}X^{j}\frac{G_{1i}%
B_{1j}-G_{1j}B_{1i}}{G_{11}}\right)  . \label{2nd term-1}%
\end{align}
Substitution of (\ref{2nd term-1}) into (\ref{action-1}) gives:%
\begin{align}
S  &  =-\sqrt{\lambda}\int d\tau\frac{d\sigma}{2\pi}\left[  p^{\alpha}\left(
\partial_{\alpha}X^{N}\frac{G_{1N}}{G_{11}}-\gamma_{\alpha\beta}%
\epsilon^{\beta\rho}\partial_{\rho}X^{N}\frac{B_{1N}}{G_{11}}\right)
-\frac{\gamma_{\alpha\beta}p^{\alpha}p^{\beta}}{2G_{11}}\right. \nonumber\\
&  +\frac{1}{2}\gamma^{\alpha\beta}\partial_{\alpha}X^{i}\partial_{\beta}%
X^{j}\left(  G_{ij}-\frac{G_{1i}G_{1j}-B_{1i}B_{1j}}{G_{11}}\right)
\nonumber\\
&  \left.  -\frac{1}{2}\epsilon^{\alpha\beta}\partial_{\alpha}X^{M}%
\partial_{\beta}X^{N}\left(  B_{MN}-\frac{G_{1M}B_{1N}-G_{1N}B_{1M}}{G_{11}%
}\right)  \right]  . \label{A9}%
\end{align}
We will use now the conservation of $p^{\alpha}$:%
\begin{equation}
\partial_{\alpha}p^{\alpha}=0, \label{conservation}%
\end{equation}
to write down the general solution to (\ref{conservation}) as:%
\begin{equation}
p^{\alpha}=\epsilon^{\alpha\beta}\partial_{\beta}\tilde{X}^{1}, \label{p-dual}%
\end{equation}
where $\tilde{X}^{1}$ is a scalar field which is the T-dual of $X^{1}$.

If we substitute for $p^{\alpha}$\ from (\ref{p-dual}) in to its definition
(\ref{p}), we find the relation:%
\begin{equation}
\epsilon^{\alpha\beta}\partial_{\beta}\tilde{X}^{1}=\gamma^{\alpha\beta
}\partial_{\beta}X^{M}G_{1M}-\epsilon^{\alpha\beta}\partial_{\beta}X^{M}%
B_{1M}. \label{relation}%
\end{equation}
Now we can derive the T-dual action by substituting for $p^{\alpha}$\ the
expression (\ref{p-dual}).

Let us consider the different terms separately.

Obviously $\left(  ij\right)  $ components remain the same since $\tilde
{X}^{i}=X^{i}$.

\noindent a) The first term in (\ref{A9}) becomes
\begin{align}
\frac{p^{\alpha}\partial_{\alpha}X^{N}G_{1N}}{G_{11}}  &  =p^{\alpha}%
\partial_{\alpha}X^{1}+p^{\alpha}\partial_{\alpha}\tilde{X}^{i}\frac{G_{1i}%
}{G_{11}}\nonumber\\
&  =p^{\alpha}\partial_{\alpha}X^{1}-\epsilon^{\alpha\beta}\partial_{\alpha
}\tilde{X}^{1}\partial_{\beta}\tilde{X}^{i}\frac{G_{1i}}{G_{11}}, \label{13}%
\end{align}
where we substitute $p^{\alpha}$ in the second term with $\epsilon
^{\alpha\beta}\partial_{\beta}\tilde{X}^{1}$.

We need also expression for $\partial_{\alpha}X^{1}$ in terms of $\tilde
{X}^{M}$. From (\ref{relation}) we have:%
\begin{equation}
\epsilon^{\alpha\beta}\partial_{\beta}\tilde{X}^{1}=\gamma^{\alpha\beta
}\partial_{\beta}X^{1}G_{11}+\gamma^{\alpha\beta}\partial_{\beta}\tilde{X}%
^{i}G_{1i}-\epsilon^{\alpha\beta}\partial_{\beta}\tilde{X}^{i}B_{1i},
\end{equation}
and therefore:%
\begin{equation}
\partial_{\alpha}X^{1}=\gamma_{\alpha\rho}\epsilon^{\rho\beta}\partial_{\beta
}\tilde{X}^{1}\frac{1}{G_{11}}+\gamma_{\alpha\rho}\epsilon^{\rho\beta}%
\partial_{\beta}\tilde{X}^{i}\frac{B_{1i}}{G_{11}}-\partial_{\alpha}\tilde
{X}^{i}\frac{G_{1i}}{G_{11}}.\label{dX1}%
\end{equation}
%or%
%\begin{equation}
%\partial_{\alpha}X^{1}=\gamma_{\alpha\rho}\epsilon^{\rho\beta}\partial_{\beta
%}\tilde{X}^{M}\frac{G_{1M}}{G_{11}}-\partial_{\alpha}\tilde{X}^{i}\frac
%{G_{1i}}{G_{11}}. \label{dX1}%
%\end{equation}
Substituting (\ref{dX1}) into (\ref{13}) we get%
\begin{equation}
p^{\alpha}\partial_{\alpha}X^{1}=\gamma^{\sigma\beta}\partial_{\sigma}%
\tilde{X}^{1}\partial_{\beta}\tilde{X}^{1}\frac{1}{G_{11}}+\gamma^{\sigma
\beta}\partial_{\sigma}\tilde{X}^{1}\partial_{\beta}\tilde{X}^{i1}\frac
{B_{1i}}{G_{11}}-\epsilon^{\sigma\alpha}\partial_{\alpha}\tilde{X}^{1}%
\partial_{\alpha}\tilde{X}^{i}\frac{G_{1i}}{G_{11}}.
\end{equation}
\noindent b) The second term in (\ref{A9}) becomes%
\begin{equation}
-p^{\alpha}\gamma_{\alpha\beta}\epsilon^{\beta\rho}\partial_{\rho}\tilde
{X}^{i}\frac{B_{1i}}{G_{11}}=-\epsilon^{\alpha\sigma}\gamma_{\alpha\beta
}\epsilon^{\beta\rho}\partial_{\rho}\tilde{X}^{i}\frac{B_{1i}}{G_{11}}%
\partial_{\sigma}\tilde{X}^{1}=\gamma^{\sigma\rho}\partial_{\sigma}\tilde
{X}^{1}\partial_{\rho}\tilde{X}^{i}\frac{B_{1i}}{G_{11}}.
\end{equation}
\noindent c) The third term in (\ref{A9}) can be written as%
\begin{equation}
-\frac{1}{2}\frac{p^{\alpha}\gamma_{\alpha\beta}p^{\beta}}{G_{11}}=-\frac
{1}{2}\epsilon^{\alpha\sigma}\frac{\gamma_{\alpha\beta}}{G_{11}}%
\epsilon^{\beta\rho}\partial_{\sigma}\tilde{X}^{1}\partial_{\rho}\tilde
{X}=\frac{1}{2}\gamma^{\sigma\rho}\partial_{\sigma}\tilde{X}^{1}\partial
_{\rho}\tilde{X}^{i}\frac{1}{G_{11}}.
\end{equation}
Summing up all the terms we derived above we find%
\begin{equation}
\frac{1}{2}\gamma^{\alpha\beta}\partial_{\alpha}\tilde{X}^{1}\partial_{\beta
}\tilde{X}^{1}\frac{1}{G_{11}}-\frac{\epsilon^{\alpha\beta}}{2}\partial
_{\alpha}\tilde{X}^{1}\partial_{\beta}\tilde{X}^{i}\frac{G_{1i}}{G_{11}}%
+\frac{1}{2}\gamma^{\alpha\beta}\partial_{\alpha}\tilde{X}^{1}\partial_{\beta
}\tilde{X}^{i}\frac{B_{1i}}{G_{11}}.
\end{equation}
All the other terms in the action remain unchanged. The final action has the
same form as (\ref{action}) but with new background fields%
\begin{equation}
S=-\frac{\sqrt{\lambda}}{2}\int d\tau\frac{d\sigma}{2\pi}\left[
\gamma^{\alpha\beta}\partial_{\alpha}\tilde{X}^{M}\partial_{\beta}\tilde
{X}^{N}\tilde{G}_{MN}-\epsilon^{\alpha\beta}\partial_{\alpha}\tilde{X}%
^{M}\partial_{\beta}\tilde{X}^{N}\tilde{B}_{MN}\right]  ,
\end{equation}
with the following transformation laws for the background fields%
\begin{align}
\tilde{G}_{11} &  =\frac{1}{G_{11}}\text{, \ \ }\tilde{G}_{ij}=G_{ij}%
-\frac{G_{1i}G_{1j}-B_{1i}B_{1j}}{G_{11}}\text{, \ \ }\tilde{G}_{1i}%
=\frac{B_{1i}}{G_{11}},\nonumber\\
\tilde{B}_{ij} &  =B_{ij}-\frac{G_{1i}B_{1j}-B_{1i}G_{1j}}{G_{11}}\text{,
\ \ }\tilde{B}_{1i}=\frac{G_{1i}}{G_{11}},
\end{align}
and the following relations between the variables%
\begin{align}
\tilde{X}^{i} &  =X^{i},\nonumber\\
\epsilon^{\alpha\beta}\partial_{\beta}\tilde{X}^{1} &  =\gamma^{\alpha\beta
}\partial_{\beta}X^{M}G_{1M}-\epsilon^{\alpha\beta}\partial_{\beta}X^{M}%
B_{1M},
\end{align}
or, equivalently%
\begin{align}
\partial_{\alpha}X^{1} &  =\gamma_{\alpha\rho}\epsilon^{\rho\beta}%
\partial_{\beta}\tilde{X}^{1}\frac{1}{G_{11}}+\gamma_{\alpha\rho}%
\epsilon^{\rho\beta}\partial_{\beta}\tilde{X}^{i}\frac{B_{1i}}{G_{11}%
}-\partial_{\alpha}\tilde{X}^{i}\frac{G_{1i}}{G_{11}}\nonumber\\
&  =\gamma_{\alpha\rho}\epsilon^{\rho\beta}\partial_{\beta}\tilde{X}^{M}%
\tilde{G}_{1M}-\partial_{\alpha}\tilde{X}^{M}\tilde{B}_{1M}.
\end{align}
These completes the derivation of the T-duality transformations.

\end{document}